\documentclass[final,1p,times]{elsarticle}
\usepackage{amssymb}
\usepackage{amsmath}
\usepackage{url}
\usepackage{amsthm}
\usepackage{graphicx}

\makeatletter
\@ifundefined{textcolor}{}
{
 \definecolor{BLACK}{gray}{0}
 \definecolor{WHITE}{gray}{1}
 \definecolor{RED}{rgb}{1,0,0}
 \definecolor{GREEN}{rgb}{0,1,0}
 \definecolor{BLUE}{rgb}{0,0,1}
 \definecolor{CYAN}{cmyk}{1,0,0,0}
 \definecolor{MAGENTA}{cmyk}{0,1,0,0}
 \definecolor{YELLOW}{cmyk}{0,0,1,0}
}

\@ifundefined{textcolor}{}
{
 \definecolor{BLACK}{gray}{0}
 \definecolor{WHITE}{gray}{1}
 \definecolor{RED}{rgb}{1,0,0}
 \definecolor{GREEN}{rgb}{0,1,0}
 \definecolor{BLUE}{rgb}{0,0,1}
 \definecolor{CYAN}{cmyk}{1,0,0,0}
\definecolor{MAGENTA}{cmyk}{0,1,0,0}
 \definecolor{YELLOW}{cmyk}{0,0,1,0}
}

\makeatother

\begin{document}
\begin{frontmatter}
\title{Estimated Age of the Universe in Fractional Cosmology}

\author[1]{Emanuel Wallison de Oliveira Costa}
\author[2]{Raheleh Jalalzadeh}
\author[1]{Pedro Felix da Silva Júnior}
\author[3,4]{Seyed Meraj Mousavi Rasouli}
\author[1]{Shahram Jalalzadeh}

\address[1]{Departamento de F\'{i}sica, Universidade Federal de Pernambuco,
Recife, PE 50670-901, Brazil}

\address[2]{Department of Physics, University of Kurdistan, Pasdaran St., Sanandaj P.O. Box 66177-15175, Iran}

\address[3]{Departamento de F\'{i}sica,
Centro de Matem\'{a}tica e Aplica\c{c}\~{o}es (CMA-UBI),
Universidade da Beira Interior,
 Rua Marqu\^{e}s d'Avila
e Bolama, 6200-001 Covilh\~{a}, Portugal.}

\address[4]{Department of Physics, Qazvin Branch, Islamic Azad University, Qazvin 341851416, Iran}


\begin{abstract}
Our proposed cosmological framework, which is based on fractional quantum cosmology, aims to address the issue of synchronicity in the age of the universe. To achieve this, we have developed a new fractional $\Lambda$CDM cosmological model. We obtained the necessary formalism by obtaining the fractional Hamiltonian constraint in a general minisuperspace. This formalism has allowed us to derive the fractional Friedmann and Raychaudhuri equations for a homogeneous and isotropic cosmology. Unlike the traditional de Sitter phase, our model exhibits a power-law accelerated expansion in the late-time universe, when vacuum energy becomes dominant. By fitting the model's parameters to cosmological observations, we determined that the fractional parameter of L\'{e}vy equals $\alpha=1.986$. Additionally, we have calculated the age of the universe to be 13.8196 Gyr. Furthermore, we have found that the ratio of the age to Hubble time from the present epoch to the distant future is finite and confined within the interval $0.9858\leq Ht<95.238$.
 \end{abstract}

\medskip

\end{frontmatter}

\section{Introduction}
\label{SecI}

Estimates for the age of the universe have been made over time. Einstein and de Sitter's model~\cite{Einstein_1932} predicted an age of the universe that was roughly $9$ Gyr, which is quite different from the age of the ancient stars. The age of ancient stars has been estimated to be larger than $12$ Gyr, as per the research by Chaboyer~\cite{Chaboyer_1995}. This discrepancy led to the development of new models that could account for the universe's observed age and expansion rate. Despite the universe experiencing deceleration for the first 9 billion years and then transitioning to acceleration through cosmic jerk for the past 5 billion years~\cite{Arturo_2016}, the present value of the dimensionless age, $H_0t_0$, where $H_0$ and $t_0$ are the current value of the Hubble parameter and the universe's age, respectively, is limited by supernova distances to be very close to 1. For example, \mbox{Tonry et al.~\cite{SupernovaSearchTeam:2003cyd}} estimated this to be $H_0t_0 = 0.96 \pm 0.04$. We seem to be experiencing a period of privilege at the moment, wherein the age of the universe in terms of Hubble time, especially in the case of a $\Lambda$CDM Universe, is extremely close to unity.
 This coincidence is known as the \emph{{synchronicity problem}} 
 of the universe's age~\cite{Arturo_2016,Juan_2020}.

Various other methods have verified this unexpected closeness to unity, which has been the subject of intense speculation and debate.
The synchronicity problem has been investigated by cosmological models with the inclusion of scalar fields~\cite{Tian_2022,Albuquerque_2021}, effective dependency between dark energy and dark matter~\cite{Kleidis_2016,Kleidis_2017,Joseph_2022,Teixeira_2022,Liu_2022}, interacting holographic dark energy~\cite{Nayak_2019,Rivera_2022,Landim_2022}, and other interpretations at the level of standard cosmology~\cite{Shimon_2022,Anari_2022}. 
From another perspective, quantum cosmology in the paradigm of the Wheeler--DeWitt (WDW) equation in the minisuperspace of a homogeneous and isotropic universe was used to propose a solution to the question in a unified way to other gaps in the $\Lambda$CDM model~\cite{Jalalzadeh_2022b}.
Proposals that modify the Friedmann--Lamaître--Robertson--Walker (FLRW)  cosmology were adopted to solve cosmological problems. They have a linear or quasi-linear evolution of the universe in common and present synchronicity $H_0t_0=1$.

Additionally, some proposals in cosmology, including the widely accepted Kolb model ($1989$) ~\cite{Kolb_1989}, explain the presence of K-matter, a unique form of exotic matter that dominates at low redshifts. K-matter has an equation of state with a value of $\omega=-1/3$, and its density decreases as the universe expands, following an inverse square relation with the scale factor. Another proposal worth mentioning is the Allen model~\cite{Allen_1999}, which incorporates a $SU(2)$ cosmological instanton in addition to the Einstein field equations. The instanton dominates during the later stages of the Universe's evolution, resulting in an approximately proportional relationship between the scale factor and cosmic time~$t$.
Moving on, Pimentel and Diaz--Rivera~\cite{Pimentel_1999} presented a solution to the synchronicity problem by considering the effect of a time-dependent cosmological constant $\Lambda$ in a family of scalar--tensor theories. Interestingly, the cosmological constant in these models exhibits an inverse square relationship with time, reaching its minimum value in the present era. However, a point of concern arises when considering the gravitational constant $G$, which decreases inversely with time and contradicts the experimental constraints on time variation in $G$. 
Moreover, Melia and Shevchuk~\cite{Melia_2012} raised an intriguing question regarding the coincidence between the cosmic time $t_0=H_0^{-1}$, describing the distance that light has traveled since the Big Bang and the gravitational horizon $R_h(t_0)$. The query was whether these two quantities were indeed equal or not. However, Lewis~\cite{Lewis_2013} criticized this model by highlighting that if the dark energy component of the universe possesses an equation of state with $\omega=-1/3$, the addition of any amount of matter will inevitably lead to a deviation from the strictly linear evolution of the $R_h=ct$ universe. This deviation would manifest the universe transitioning to a matter-dominated phase during its earliest epochs.

The primary aim of this study is to establish an appropriate fractional cosmological framework to examine the synchronicity issue, among other enigmas linked to the $\Lambda$CDM model. Our investigation endeavors to tackle the synchronicity problem through the fractional quantum cosmology (FQC) proposal~\cite{Moniz_2020,Moniz_2021,Jalalzadeh_2021,Jalalzadeh_2022,Rasouli:2022bug,Jalalzadeh_2022b,Jalalzadeh_2022c}. The basic and preliminary methodology of the FQC is to combine fractional quantum mechanics (FQM) ~\cite{Laskin_2000,Laskin_2002,Laskin_2010} and quantum cosmology (QC)~\cite{Calcagni_2017} to achieve the corresponding cosmological consequences (see~\cite{Jalalzadeh_2022c} for a detailed discussion).
However, it should be mentioned that in this study, we only consider the (semi-)classical limit of this theory; to see other semiclassical models established applying various gravitational theories as underlying frameworks, see, for instance,~\cite{Rasouli:2011zkm,Rasouli:2013sda,Jalalzadeh:2014jea,Rasouli:2014dba,Rasouli:2016xdo,Rasouli:2016syh,Rasouli:2018nwi,Rasouli:2018lny,Rasouli:2022hnp}. Several studies have put forth the idea of fractional classical cosmology (FCC)~\cite{Shchigolev_2011,Gabriele_2021,Garcia_2022}, since the application of fractional calculus in cosmology is now producing encouraging and promising results~\cite{Leon_2023}.

There have been numerous instances in the fields of gravity and cosmology where fractional calculus has been employed, showcasing the broad and significant applications of this mathematical tool. The current research in this domain is thriving, with scientists actively exploring the potential of fractional calculus. Its effectiveness as a valuable instrument becomes apparent when confronting a wide range of problems associated with gravitational forces and cosmological models, as evidenced by the wealth of studies conducted~\citep{Leon_2023,Garcia_2022,Shchigolev:2012rp,Shchigolev:2013jq,Calcagni:2013yqa,Shchigolev:2015rei,Calcagni:2016azd,Shchigolev:2021lbm,Calcagni:2020ads,Calcagni:2021ipd,Calcagni:2021aap,Gonzalez:2023who,Socorro:2023ztq}. Moreover, fractional calculus finds application in various other areas, such as the study of the stochastic gravitational-wave background in quantum gravity~\citep{Calcagni:2020tvw}, the investigation of gravitational-wave luminosity distance~\citep{Calcagni:2019ngc}, the exploration of inflation and CMB spectrum~\citep{Rasouli:2022bug,Calcagni:2017via,Calcagni:2016ofu}, the development of fractional action cosmology~\citep{El-Nabulsi:2012wpc,El-Nabulsi:2015szp,Jamil:2011uj}, the analysis of the fractional geodesic equation and discrete gravity~\cite{El-Nabulsi:2013mma}, and the examination of nonminimal coupling and chaotic inflation~\cite{El-Nabulsi:2013mwa}. Additionally, fractional calculus has been utilized in investigating phantom cosmology with conformal coupling~\cite{Rami:2015kha}, Ornstein--Uhlenbeck-like fractional differential equations in cosmology~\cite{El-Nabulsi:2016dsj}, fractional action cosmology with a variable-order parameter~\cite{El-Nabulsi:2017vmp}, and wormholes in fractional action cosmology~\citep{El-Nabulsi:2017jss}. 
Notably, new metrics have been considered~\citep{El-Nabulsi:2017rdu}, while the application of fractional calculus has also extended to the study of dark energy models in emergent, logamediate, and intermediate scenarios of the universe~\cite{Debnath2012,Debnath2013}. As an example, the work of~\cite{Shchigolev:2015rei,Shchigolev:2021lbm} yielded a value of $\alpha=0.926$ for the order of the Riemann--Liouville fractional integral, providing further insights into the intricacies of the field.

In~\cite{Shchigolev_2011,Shchigolev:2012rp,Shchigolev:2013jq}, several exact solutions were obtained for cosmological models, which, due to the fractal nature of space--time, exhibit a significant deviation from the standard model~\cite{Calcagni:2010bj,Calcagni:2009kc}. The investigation carried out in ref.~\cite{Jalalzadeh_2022} delves into the interval $1\leq\alpha <2$ and employs Riesz's fractional derivative to derive the nonboundary and tunneling wave functions for a closed de Sitter geometry. Furthermore, ref.~\cite{Rasouli:2022bug} explores the preinflation epoch within the framework of fractional quantum cosmology. With regard to the thermodynamics of fractional BHs, this subject has been thoroughly examined in ref.~\cite{Jalalzadeh_2022}, and the findings from this research have been effectively utilized to obtain the Friedmann equations in the context of emergent gravity in ref.~\cite{Junior:2023fwb}. It is worth mentioning that fractional calculus has also been employed to modify the Friedmann and Raychaudhuri equations, thereby enabling an investigation into the dynamics of the universe without the need for cold dark matter (CDM) or dark energy~\cite{Barrientos:2020kfp}. 
In addition, an alternative approach involves the utilization of fractional calculus to determine the value of the cosmological constant, which necessitates a restructuring due to the well-recognized ultraviolet divergence in conventional quantum field theory~\cite{Landim:2021www,Calcagni_2021,Landim:2021ial}. To expand the scope of this fractional approach,~\cite{Giusti:2020rul,Torres:2020xkw} explore modified Newtonian dynamics (MOND) and quantum cosmology. Lastly, it is crucial to note that there exists a multitude of definitions for fractional derivatives and fractional integrals, including those proposed by Riemann--Liouville, Caputo, Riesz, Hadamard, Marchand, and Griinwald--Letnikov, as well as more recent formulations (as comprehensively documented in~\cite{book1:2006} and~\cite{book2:1999}, along with their respective references). Despite the extensive research conducted on these operators, they do not always exhibit the conventional properties associated with function differentiation, such as Leibniz's rule, the chain rule, and the semigroup property~\cite{book2:1999,book1:2006}.

This paper is organized as follows: We begin in Section \ref{section3} by introducing fractional quantum cosmology, which allows us to define the fractional extension of the ADM Hamiltonian and Hamiltonian constraint within the minisuperspace approximation of gravity. Next, in Section \ref{section33}, we apply the general formalism obtained in Section \ref{section33} to the FLRW cosmology. This leads to deriving the fractional extensions of the Friedmann and Raychaudhuri equations. We then proceed with a thorough discussion of various fractional cosmological parameters, including the Hubble and deceleration parameters. 
In Section \ref{section5}, we briefly review the available cosmological datasets. Afterwards, we consider the observational datasets, such as SNIa, CMB, BAO, BBN, and OHD, to estimate both the universe's age and the model's free parameters. Section \ref{section4} focuses on the perspective of fractional cosmology regarding the synchronicity problem. Finally, in Section \ref{conclusions}, we summarize the results and offer additional discussions. 

\section{Fractional Quantum Cosmology} \label{section3}
\indent

This section briefly explains the FQC framework. To develop a fractional WDW equation, we first briefly describe how to construct a fractional Schr\"{o}dinger
equation (SE).

In two seminal papers, Laskin~\cite{Laskin_2000,Laskin:1999tf} introduced a groundbreaking concept known as space fractional quantum mechanics (SFQM) through the application of a purely physics-based approach. Laskin's revolutionary work was built upon the solid foundation of nonrelativistic quantum mechanics as formulated by Feynman and Hibbs, which extensively utilized a path integral representation employing Brownian paths. However, Laskin's innovative contribution to the field was to replace these conventional Brownian paths with paths characterized by L\'{e}vy flights, thus giving rise to the fascinating realm of SFQM and expanding our understanding of quantum phenomena in space. 

In the framework of SFQM, one can initiate the analysis by considering the conventional time-dependent {SE, which is expressed as follows:} 
\begin{equation} \label{SE1}
i\hbar\frac{\partial\psi(\textbf{r},t)}{\partial{t}}=-\frac{\hbar^2}{2m}\Delta\psi(\textbf{r},t)+V(\textbf{r},t)\psi(\textbf{r},t).
\end{equation}

This equation serves as a starting point for further exploration, wherein one can delve into its fractional counterpart, as proposed by Laskin~\cite{Laskin_2002}. The fractional SE is obtained by replacing the traditional Laplace operator $\Delta$ with the Riesz fractional derivative $(-\hbar^2\Delta)^{\alpha/2}$, which can be expressed as
\begin{equation} \label{operator1}
-\frac{\hbar^2}{2m}\Delta\, \rightarrow \, D_\alpha\left(-\hbar^2\Delta\right)^{\alpha/2}.
\end{equation}

This replacement paves the way for the exploration of the space-fractional SE: 
\begin{equation}
    i\hbar \frac{\partial \psi(\mathbf{r},t)}{\partial  t}
= \hat{H}_\alpha(\hat{\mathbf{p}}, \hat{\mathbf{r}}) \psi(\mathbf{r},t)
=
D_\alpha
(-\hbar^2 \Delta)^{\alpha/2} \psi(\mathbf{r},t)
+ V(\mathbf{r},t) \psi(\mathbf{r},t),~~~~1 < \alpha \leq 2,
\label{7}
\end{equation}
where $\alpha\in(1,2]$ is the L\'evy index, and $D_\alpha$ is, according to Laskin, a generalized ``fractional quantum
diffusion coefficient'' of dimension $\text{J}^{1-\alpha}\text{m}^\alpha \text{s}^{-\alpha}$. As mentioned earlier, $(-\hbar^2 \Delta)^{\alpha/2}$ is a generalization of the fractional Riesz derivative, known as the fractional Laplacian~\cite{Pozrikidis_2018}:
\begin{equation}
(-\hbar^2 \Delta)^{\alpha/2} \psi(\mathbf{r},t)
=
\frac{1}{(2\pi\hbar)^3}
\int d^3 p e^{i\frac{\mathbf{p}\cdot \mathbf{r}}{\hbar}}
|\mathbf{p}|^\alpha
\varphi(\mathbf{p},t),
\label{8}
\end{equation}
which, by means of Fourier transforms, relates
$\psi(\mathbf{r}, t)$ and   $\varphi(\mathbf{p}, t)$.
For the special case when $\alpha = 2$, $D_\alpha$ reduces to $D_2
= 1/2m$, where $m$ is the particle's mass.

The Riesz fractional derivative (the fractional Laplacian), in general, characterizes a unique diffusion process that arises from the random displacements of individuals who can move to nearby or neighboring locations and even venture to distant sites by means of Lévy flights. Both literal and conceptual flights have been observed or claimed to occur in various scenarios, such as turbulent fluid motion and the transportation of materials in fractured media. In the field of mechanics, the fractional Laplacian describes the motion of a chain or group of particles interconnected by elastic springs that connect not only to their immediate neighbors but also to all other particles. The strength of the springs decreases as the particles move apart, while the particle arrangement can take on either a regular or fractal pattern. In a more abstract sense, the fractional Laplacian represents the influence of a nonlocal process on a conservation law, which is not only influenced by local conditions but also by the overall state of a field of interest at a specific moment in time. We suggest that the reader consults Pozrikidis's exceptionally well-written and comprehensive book~\cite{Pozrikidis_2018} in order to obtain detailed explanations and develop a deeper understanding of fractional Laplacian.

On the other hand, as demonstrated by Laskin in his work~\cite{doi:10.1142/10541}, the evolution of the fractional wave function (\ref{7}) can be effectively described by a kernel. This kernel is expressed as an integral equation, which can be written as follows:
\begin{equation} \label{Kernel1} \psi(\mathbf{r}_b,t_b)=\int d^3r_a K_L(\mathbf{r}_b,t_b|\mathbf{r}_a,t_a)\psi(\mathbf{r}_a,t_a), \end{equation}

In this equation, $\psi(\mathbf{r}_a,t_a)$ represents the wave function of the initial state, while $\psi(\mathbf{r}_b,t_b)$ represents the wave function of the final state. The kernel itself, denoted as $K_L(\mathbf{r}_b,t_b|\mathbf{r}_a,t_a)$, is given by the following expression:
\begin{equation} \label{Kernel2} K_L(\mathbf{r}_b,t_b|\mathbf{r}_a,t_a)=\int_{\mathbf{r}_a}^{\mathbf{r}_b}D\mathbf{r}(\tau)\int D\mathbf{p}(\tau)e^{\frac{i}{\hbar}\int_{t_a}^{t_b}(\mathbf{p}(\tau)\cdot\dot{\mathbf{r}}(\tau)-H_\alpha)d\tau}. \end{equation}

Here, it is important to note that $H_\alpha(\mathbf{r},\mathbf{p},t)$ serves as the fractional extension for the Hamiltonian of the system under consideration. This Hamiltonian can be mathematically expressed as
\begin{equation} \label{fracclasshamil} H_\alpha(\textbf{p},\textbf{r})={D_\alpha}|\textbf{p}|^\alpha+V(\textbf{r},t), \,\,\,\,\,\,\, 1<\alpha\leq{2}. \end{equation}

It is worth mentioning that in the particular scenario where $\alpha$ takes a value of 2 and $D_\alpha$ is equal to $1/{2m}$, the equation \eqref{fracclasshamil} simplifies to the standard Hamiltonian, which is defined as $H(\textbf{p},\textbf{r},t)=\textbf{p}^2/2m+V(\textbf{r},t)$.

We can further expand the aforementioned procedure in order to acquire fractional quantum cosmology.
In cosmology, we assume that the space--time metric and matter field(s) exhibit homogeneity and isotropy. Consequently, the lapse function is considered to be homogeneous, denoted as $N = N(t)$, while the shift is assigned a value of zero, $N^i= 0$. As a result, the space--time line element reduces to
\begin{equation}
    \label{SS1}
    ds^2=-N(t)^2dt^2+h_{ij}(\mathbf{x},t)dx^idx^j.
\end{equation}

Inserting the restricted form of the line element (\ref{SS1}) into the ADM action functional, one generally obtains the following reduced form of the action~\cite{Jalalzadeh_2022c}: 
\begin{equation}
    \label{SS2}
  S[q^\mu(t),N(t)]=\int\left\{\frac{1}{2N(t)}f_{\mu\nu}(q)\dot q^\mu \dot q^\nu -N(t)U(q)\right\}dt,~~~~~\mu,\nu=1,2,...,n.
\end{equation}

In this equation, $f_{\mu\nu}$ denotes the reduced DeWitt metric of $n$-dimensional minisuperspace with signature $(-,+,...,+)$. Also, $U(q)$, the potential, includes the Ricci scalar of the three-dimensional $t=const.$ hypersurfaces and the potential of matter fields. 
Generally, the minisuperspace coordinates $q^\mu$ may include matter fields and three-metric components. Variation in the action with respect to $q^{ A}$  yields the Euler--Lagrange equations
\begin{eqnarray}\label{reducedgeodesic}
\frac{1}{N}\frac{d}{dt}\Big(\frac{\dot q^{\mu}}{N}\Big)+\frac{1}{N^2}\Big\{^{\mu}_{\alpha\beta}\Big\}\dot q^{\alpha}\dot q^{\beta}=-f^{\mu\nu}\partial_{\nu}U(q),
\end{eqnarray}
where  $\Big\{^{\mu}_{\alpha\beta}\Big\}$ are the connection coefficients (Christoffel symbols) determined from the metric of the minisuperspace $f_{\mu\nu}$. Also, in a variation with respect to the lapse function, $N$, one obtains the constraint 
\begin{equation}
    \label{Hcons}
    \frac{1}{2N}f_{\mu\nu}\dot q^\mu\dot q^\nu+NU(q)=0.
\end{equation}

To ensure consistency, it is necessary for Equations (\ref{reducedgeodesic}) and (\ref{Hcons}) to be equal to the $00$ and $ij$ components of the complete Einstein equations, respectively.

To define the Hamiltonian, we first need to define canonical momenta in a usual way, which is
\begin{equation}
    \label{moment}
    \Pi_\mu=\frac{\partial L}{\partial \dot q^\mu}=\frac{1}{N}f_{\mu\nu}\dot q^\nu.
\end{equation}

This leads us to the canonical ADM Hamiltonian in the minisuperspace
\begin{equation}
    \label{Ham111}
    H_\text{ADM}=N\left(\frac{1}{2}f^{\mu\nu}(q)\Pi_\mu\Pi_\nu+U(q) \right)\equiv N\mathcal H,
\end{equation}
where $f^{\mu\nu}$ is the inverse metric of the minisuperspace.
The lapse function $N$ serves as a Lagrange multiplier with the purpose of ensuring the satisfaction of the Hamiltonian constraint
\begin{equation}
    \label{Ham222}
    \mathcal H(q^\mu,\Pi_\mu)=\frac{1}{2}f^{\mu\nu}\Pi_\mu\Pi_\nu+U(q)=0.
\end{equation}

In the coordinate representation of canonical quantization, the process involves the incorporation of a wave function, denoted as $\Psi(q)$, which is independent of time and necessitates the condition that it is rendered null by the operator associated with the aforementioned Hamiltonian constraint. As a result, the WDW equation is obtained:
\begin{equation}
    \label{Ham333}
    \left\{-\frac{1}{2}\Delta +\zeta\mathcal R+U(q)  \right\}\Psi(q)=0,
\end{equation}
where $\Delta=\frac{1}{\sqrt{-f}}
\partial_\alpha(\sqrt{-f}f^{\alpha\beta}\partial_\beta)$ is the d'Alembertian, $\mathcal R$ is the curvature of the minisuperspace metric $f_{\mu\nu}$,
and $\zeta$ is an arbitrary constant~\cite{Jalalzadeh_2022c}.
It is important to note that the minisuperspace metric's dependence on the local coordinates of the minisuperspace leads to a nontrivial operator ordering issue in the resulting WDW equation. However, this issue can be somewhat addressed by ensuring that the quantization process is covariant in minisuperspace. This means that it should not be affected by general coordinate transformations in the minisuperspace. The factor ordering procedure is the origin of the minisuperspace curvature term, denoted by $\mathcal R$.

To acquire the fractional equivalent of the WDW equation \eqref{Ham333}, it is necessary to substitute the typical d'Alembertian operator with the fractional Riesz--d'Alembertian operator given by
\cite{Rie,Tarasov:2018zjg}
\begin{equation}\label{Riesz-d'Alembertian}
(-\nabla)^{\frac{\alpha}{2}} \Psi(q^\alpha)
={\mathcal F}^{-1}\Big(|{\mathbf{\Pi}}|^\alpha{\mathcal F}\Psi({\mathbf{\Pi}})\Big),
\end{equation}
where $|{\mathbf{\Pi}}|=\sqrt{f^{\mu\nu}\Pi_\mu \Pi_\nu}$, and ${\mathcal F}$ represents a Fourier transformation. Therefore, the fractional equivalent of the WDW equation (\ref{Ham333}) can be expressed as~\cite{Moniz_2020,Moniz_2021,Jalalzadeh_2021,Jalalzadeh_2022,Jalalzadeh_2022c}
\begin{equation}\label{WDWsh}
    \left\{\frac{D_\alpha(m_\text{P})}{2}(-\nabla)^\frac{\alpha}{2}+\zeta\mathcal R+U(q)\right\}\Psi(q^\nu)=0,
\end{equation}
where $D_\alpha(m_\text{P})$ is the coefficient, depending on the Planck mass $m_\text{P}=1/\sqrt{G}$, which reduces to unity at $\alpha=2$ limit, and the value of Lévy's fractional parameter, denoted as $\alpha$, is assigned the range of values  $1 < \alpha \leq 2$.

By following the methodology described by Laskin~\cite{doi:10.1142/10541}, it becomes feasible to illustrate the comprehensive procedure for tje fractional path integral quantization of constrained systems, as expounded upon by Balatin, Fradkin, and Vilkovisky (for further elaboration, please refer to~\cite{Batalin:1977pb}). This approach culminates in the derivation of the resulting \textit{{propagation amplitude}}
  which arises when this technique is applied to the minisuperspace model that is precisely defined by the Hamiltonian (\ref{Ham222})
\begin{equation}
    \label{Vilkov}
    \begin{split}
    G_\alpha(\mathbf{q}_2,\mathbf{q}_1)&=\int dN(t_2-t_1)\int Dq^\mu D\Pi_\mu\exp\left\{i\int_{t_1}^{t_2}(\dot q^\nu\Pi_\nu-N{\mathcal H}_\alpha)dt\right\}\\
    &=\int dT\int Dq^\mu D\Pi_\mu\exp\left\{i\int_{0}^{T}(\dot q^\nu\Pi_\nu-N{\mathcal H}_\alpha)dt\right\}\\
    &=\int K_L(\mathbf{q}_2,T|\mathbf{q}_1,0)dT,
    \end{split}
\end{equation}
where $\mathcal H_\alpha(q^\mu,\Pi_\mu)$ is the fractional Hamiltonian constraint, counterpart to (\ref{Ham222}), defined by
\begin{equation}
    \label{Ha444}
    \mathcal H_\alpha(q^\mu,\Pi_\mu)=\frac{D_\alpha(m_\text{P})}{2}|f^{\mu\nu}\Pi_\mu \Pi_\nu|^\frac{\alpha}{2}+U(q)=0.
\end{equation}

This equation gives us the fractional extension of Equation (\ref{Hcons}). Also, one can easily obtain the fractional Euler--Lagrange counterpart of (\ref{reducedgeodesic}) by Hamilton's equations
\begin{equation}
    \label{FHamil}
    \dot q^\mu=\frac{\partial H^{(\alpha)}_\text{ADM}}{\partial \Pi_\mu},~~~~~\dot \Pi_\mu=-\frac{\partial H^{(\alpha)}_\text{ADM}}{\partial q^\mu},
\end{equation}
where $H^{(\alpha)}_\text{ADM}$ is the fractional ADM Hamiltonian defined by
\begin{equation}
    \label{ADMf}
    H^{(\alpha)}_\text{ADM}=N\left(\frac{D_\alpha(m_\text{P})}{2}\left| f^{\mu\nu}\Pi_\mu \Pi_\nu\right|^\frac{\alpha}{2}+U(q) \right).
\end{equation}

In the following section, we derive the fractional $\Lambda$CDM cosmology using Equations \eqref{Ha444} and \eqref{FHamil}.

\section{Fractional FLRW Cosmology} \label{section33}

Let us commence our discussion by considering the FLRW metric, which serves as the underlying geometric structure
\begin{equation} \label{line1}
ds^2=-N^2(t){\rm{d}}t^2+a^2(t)\left\{\frac{{{{\rm\,d}r^2}}}{{1-kr^2}}+r^2{\rm{d}}\Omega_2^2\right\}.
\end{equation}

 In this metric, we can observe various components contributing to the overall structure. Firstly, the term $N(t)$ represents the lapse function, which characterizes the rate at which time elapses within the cosmic framework. Secondly, the variable $t$ represents the cosmic time, enabling us to measure and understand the progression of events within the universe. Furthermore, the term ${\rm{d}}\Omega_2^2$ stands for the standard line element for $\mathbb S^2$, encompassing both the angular coordinates $\theta$ and $\phi$. Lastly, the parameter $k$ assumes values of $-1, 0,$ and $1$, signifying the sign of spatial curvature. These values correspond to an open, flat, and closed universe, respectively.
 
The functional, which encompasses both gravitational and matter components, with the matter field being treated as a perfect fluid~\cite{hawking_ellis_1973,Fathi:2017pjm,Rashki:2016udu,Jalalzadeh:2014jka,Fathi:2016lws,Jalalzadeh:2014jea,Rashki:2014noa}, can be expressed as follows:
\begin{equation}
S=\frac{1}{16\pi\,G}\left[\int_{\mathcal{M}}{\sqrt{-g}R\, \rm {d}^4x}+\int_{\partial\mathcal{M}}{2K\sqrt{h}}{\rm{d}}^3{x}\right]-\int_{\mathcal{M}}\sqrt{-g}\rho{\rm{d}}^4x,\label{action1}
\end{equation}
where $K$ denotes the trace of the extrinsic curvature of the spacelike submanifolds,  $(\Sigma,h)$, of the space--time, $(\mathcal M,g)$, in which $\mathcal M=\mathbb R\times \Sigma$ and $\mathcal M$ represents the manifold of the space--time with boundary $\partial M$.  In addition, $\rho$ is the total energy density of the cosmic fluid with components
 \begin{equation}
     \label{Fluid}
\rho=\sum_i\rho^{(i)}=\rho^{(r)}+\rho^{(m)}+\rho^{(\Lambda)},
 \end{equation}
 where $\rho^{(r)}$, $\rho^{(m)}$, and $\rho^{(\Lambda)}=\frac{\Lambda}{8\pi G}$ are the energy density of the radiation, matter density of the cosmic dust (constitute of baryonic and cold matter), and the energy density of cosmological constant, respectively. 
 
The act of introducing the homogeneous and isotropic metric \eqref{line1} into the action functional \eqref{action1} results in a simplified  Arnowitt--Deser--Misner (ADM) form as follows~\cite{Jalalzadeh_2022c}:
\begin{equation}
S_{ADM}=\int_{t_i}^{t_f}{\rm\,d}t\left[{\frac{3\mathcal{V}_k}{8\pi\,G}\left(kNa-\frac{a\,\dot{a}^2}{N}\right)-{\mathcal{V}_kNa^3\rho(t)}}\right], \label{action2}
\end{equation}
\noindent where
\begin{equation}
    \mathcal{V}_k=\int_\Sigma{{\rm{d}}r{\rm{d}}\theta{\rm{d}}\phi\frac{r^2\sin{\theta}}{\sqrt{1-kr^2}}},
\end{equation}
 is the spatial volume of the spacelike hypersurfaces $(\Sigma,h)$.

To ensure that the spacelike sections, represented by $\Sigma$, have a finite volume, it is assumed that they are both compact and without boundary. In the scientific community, it is widely accepted that any compact Riemannian three-manifold with constant curvature is homeomorphic to $\tilde\Sigma/\Gamma$, where $\Gamma$ represents the group of covering transformations, and $\tilde\Sigma$ represents the universal covering space. The shape of $\tilde\Sigma$ can take on different forms depending on the sign of the spatial curvature $k$ (where $k$ can be 0, 1, or -1), such as $\mathbb R^3$ (representing three-dimensional Euclidean space), $\mathbb S^3$ (representing a 3-sphere), or $\mathbb H^3$ (representing three-dimensional hyperbolic space)~\cite{Jalalzadeh_2022c}.

Using this particular approach, the action mentioned above is clearly defined and has a finite 3-volume, which is an important characteristic. The formula for accurately determining the volume of spherical 3-spaces is given by $\mathcal V_{k=1}=2\pi^2/|\Gamma|$, where $|\Gamma|$ represents the order of the group $\Gamma$. It is worth noting that when the 3-manifold is topologically complex, the order of the group $\Gamma$ tends to be large, resulting in a relatively small corresponding volume. Additionally, it is important to mention that there is no lower bound on the volume since the group $\Gamma$ can have an arbitrarily large number of elements according to~\cite{Jalalzadeh_2022c}.

Ongoing research and exploration within the scientific community are still being carried out to determine the allowed topologies of $\mathbb H^3/\Gamma$. However, Thurston and Jørgensen have made significant contributions in this field by demonstrating the existence of the smallest volume for compact orientable hyperbolic 3-manifolds, as discussed in~\cite{Benedetti1992}. This achievement can be attained through a finite number of manifolds, with the prime candidate being the Weeks--Matveev--Fomenko manifold. Its volume is approximately 0.942707. The research conducted by Thurston and Jørgensen has greatly enhanced our understanding of the intricate and fascinating properties of hyperbolic 3-manifolds and their volumes.

The Lagrangian that corresponds to the aforementioned action (\ref{action2}) is
\begin{equation} \label{lagran1}
L_{ADM}=\frac{3\mathcal{V}_k}{8\pi\,G}\left(kNa-\frac{a\,\dot{a}^2}{N}\right)-{\mathcal{V}_kNa^3\rho(t)}.
\end{equation}

As a result, the corresponding ADM Hamiltonian is given by
\begin{equation} \label{Hami2}
H_{ADM}=N\left[\frac{2\pi\,G\,\Pi^2}{3\mathcal{V}_k\,a}+\frac{3\mathcal{V}_k\,k\,a}{8\pi\,G}-\mathcal{V}_k\,a^3\,\rho\right],
\end{equation}
where
\begin{equation}
    \Pi=-\frac{3\mathcal V_ka\dot a}{4\pi GN},
\end{equation}
is the conjugate momenta of the scale factor.
The ability to select the lapse function provides us with gauge freedom, resulting in the super-Hamiltonian constraint
\begin{equation} \label{Hami23}
\mathcal{H}=\frac{2\pi G\Pi^2}{3\mathcal{V}_k\,a}+\frac{3\mathcal{V}_k\,k\,a}{8\pi\,G}-\mathcal{V}_k\,a^3\,\rho=0.
\end{equation}

The minisuperspace of the FLRW cosmology, corresponding to the Hamiltonian constraint (\ref{Hami23}), is unidimensional with $f^{\mu\nu}\Pi_\mu\Pi_\nu=\Pi^2/a$. Regarding dimensions of $[\Pi^2/a]=1/(\text{length})^3$ and $[\mathcal H]=1/(\text{length})$, one can easily  verify that $D_\alpha(m_\text{P})=m_\text{P}^{3(2-\alpha)}$. Therefore, the fractional extension of the ADM Hamiltonian (\ref{Hami2}) and Hamiltonian constraints (\ref{Hami23}) are given by
\begin{equation}
    \label{FADM}
    H^{(\alpha)}_\text{ADM}=N\left[\frac{2\pi l_\text{P}^{3\alpha-4}}{3\mathcal{V}_k}\left|\frac{\Pi^2}{a}\right|^\frac{\alpha}{2}+\frac{3\mathcal{V}_k\,k\,a}{8\pi\,G}-\mathcal{V}_k\,a^3\,\rho\right],
\end{equation}
\begin{equation}
    \label{FH21}
    \frac{2\pi l_\text{P}^{3\alpha-4}}{3\mathcal{V}_k}\left|\frac{\Pi^2}{a}\right|^\frac{\alpha}{2}+\frac{3\mathcal{V}_k\,k\,a}{8\pi\,G}-\mathcal{V}_k\,a^3\,\rho=0,
\end{equation}
where $l_\text{P}=\sqrt{G}$ is the Planck length.

In the comoving gauge $N=1$, using the Hamilton's equation, $\dot a=\partial H^{(\alpha)}_\text{ADM}/\partial{\Pi}$, and rescaling the cosmic time by
\begin{eqnarray}
    t\rightarrow \frac{\alpha}{4^\frac{\alpha-1}{\alpha}}\left(\frac{3\mathcal V_k}{2\pi}\right)^\frac{\alpha-2}{\alpha}t,
\end{eqnarray}
lead us to the following fractional extension of the Friedmann equation
\begin{equation}
H^\frac{2}{3-D}=\frac{1}{\left(l_\text{P}^{2}a^3\right)^\frac{D-2}{3-D}}\left[\frac{8\pi G}{3}\rho-\frac{k}{a^{2}} \right],
  \label{FracFriedmann1}
\end{equation}
where $\rho$ is given by relation (\ref{Fluid}), and the parameter $D$ is related to the  L\'evy fractional parameter $\alpha$ by
\begin{equation}
    \label{dimension}
    D=\frac{2}{\alpha}+1,~~~~~~2\leq D<3.
\end{equation}

Substituting $\alpha=2$ (or equivalently $D=2$) into Equation \eqref{FracFriedmann1}, we obtain the standard counterpart, i.e., \eqref{Frrr111}.
We should note that there are several variants of the fractional Friedmann Equation in the literature; see, for instance,~\cite{Shchigolev_2011,Gabriele_2021}.

Similar to the standard model of cosmology, one can define the fractional extension of the density parameters $\{\Omega^{(i)}=\frac{8\pi G\rho^{(i)}}{3H^2}, \Omega^{(k)}=-\frac{k}{a^2H^2}\}$ by
\begin{equation}
    \label{demsity33}
    \tilde\Omega^\text{(i)}=\frac{1}{\left(l_\text{P}^{2}a^3\right)^\frac{D-2}{3-D}}\frac{8\pi G\rho^{(i)}}{3H^\frac{2}{3-D}},~~~~~\tilde\Omega^{(k)}=-\frac{1}{\left(l_\text{P}^{2}a^3\right)^\frac{D-2}{3-D}}\frac{k}{a^2H^\frac{2}{3-D}}.
\end{equation}

It should be noted that the density parameters mentioned above follow the standard definitions in $\Lambda$CDM when $D=2$.
As a result, the fractional Friedmann (\ref{FracFriedmann1}) takes the following standard form
\begin{equation}
    \label{FF11}
\tilde\Omega^{(m)}+\tilde\Omega^{(rad)}+\tilde\Omega^{(\Lambda)}+\tilde\Omega^{(k)}=1,
\end{equation}
which is also known as the closure relation.  

The Raychaudhuri equation, also known as the second Friedmann equation, can be derived by taking the time derivative of the Friedmann Equation (\ref{FracFriedmann1}) and combining it with the continuity equation. In fractional cosmology, the energy--momentum tensor of a perfect fluid is subject to the same covariant conservation equation as in standard cosmology. This equation can be expressed as follows:
\begin{equation}
    \label{covar}
    \dot\rho^{(i)}+3H\Big(\rho^{(i)}+p^{(i)}\Big)=0,
\end{equation}
where $p^{(i)}=\omega_i\rho^{(i)}$.

These two lead us to the fractional Raychaudhuri equation
\begin{equation}
    \label{Raychaudhuri11}
    \frac{\ddot a}{a}=-\frac{4\pi(3-D)G}{3(H^2l_\text{P}^2a^3)^{\left(\frac{D-2}{3-D} \right)}}(\rho+3p)-\frac{D-2}{2}H^2.
\end{equation}

In addition, the fractional Friedmann Equation (\ref{FracFriedmann1}), utilizing the continuity Equation~(\ref{covar}), can be rewritten as
\begin{equation}
    \label{Frrr}
    \left(\frac{H}{H_0}\right)^\frac{2}{3-D}=\frac{1}{a^{3\left(\frac{D-2}{3-D}\right)}}\left\{\sum_i\frac{\tilde \Omega^{(i)}_0}{a^{3(1+\omega_i)}}+\frac{\tilde\Omega^{(k)}_0}{a^{2}}\right\},
\end{equation}
where $\tilde\Omega^{(i)}_0$ and $\tilde\Omega^{(k)}_0$ denote the density parameters at the present epoch, $(a=1)$.

In the same method, one can obtain the deceleration parameter $q$. By taking a time derivative of the Friedmann equation (\ref{FracFriedmann1}) and employing the continuity equation and fractional density parameters (\ref{demsity33}), we find 
\begin{equation}
    \label{decel}
    \begin{split}
    q&=-\frac{\ddot a}{aH^2}=\frac{3-D}{2}\sum_i(1+3\omega_i)\tilde\Omega^{(i)}+\frac{D-2}{2}\\
    &=\frac{3-D}{2\left(\frac{H}{H_0}\right)^\frac{2}{3-D}a^{3\left(\frac{D-2}{3-D} \right)}}\sum_i(1+3\omega_i)(1+z)^{3(1+\omega_i)}\tilde\Omega^{(i)}_0+\frac{D-2}{2},
    \end{split}
\end{equation}
where, in the second equality, we used the definition of the redshift $1+z=1/a$.

\begin{figure}[!h]
\centering
 \includegraphics[width=8cm]{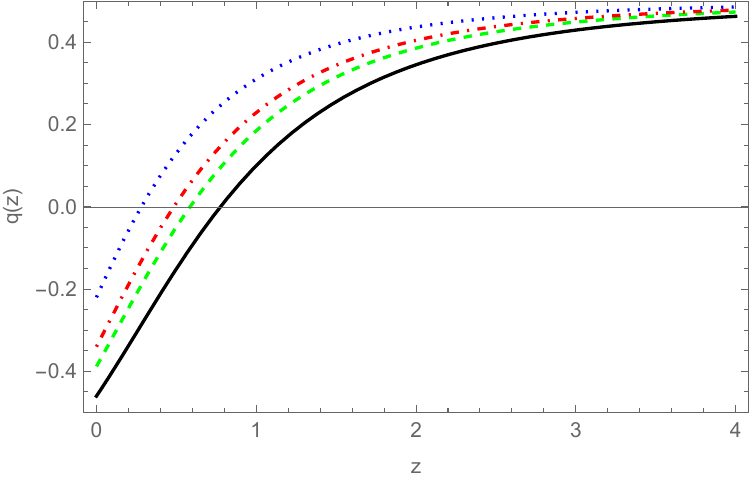}
  \caption{\small The evolution of the deceleration parameter, $q$, as a function of redshift, $z$, for $D=2.2$ and $\tilde\Omega^{(m)}_0=0.2$ (solid line), $\tilde\Omega^{(m)}_0=0.26$ (dashed), $\tilde\Omega^{(m)}_0=0.3$ (dot-dashed), and $\tilde\Omega^{(m)}_0=0.4$ (doted).}\label{FigS1}
\end{figure}

\begin{figure}[!h]
\centering
 \includegraphics[width=8cm]{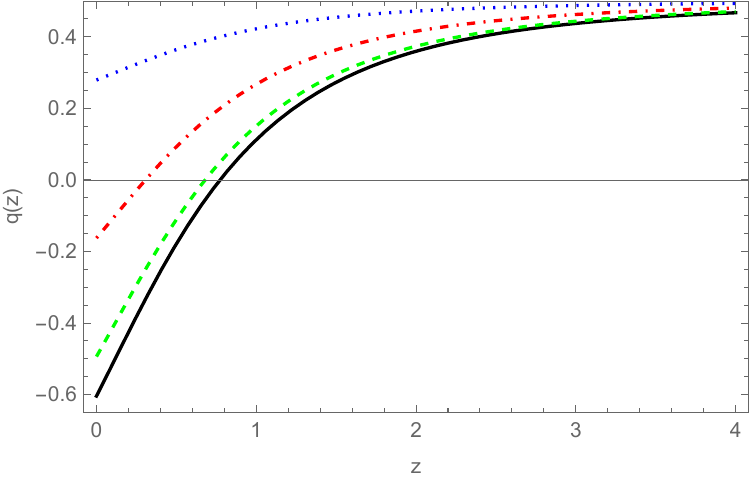}
  \caption{\small The evolution of the deceleration parameter, $q$, as a function of redshift, $z$, for $\tilde\Omega^{(m)}_0=0.264$ and $D=2$ (solid line), $D=2.1$ (dashed), $D=2.4$ (dot-dashed), and $D=2.8$ (doted).}\label{FigS2}
\end{figure}

To observe the overall impact of the Lévy fractional parameter on cosmic evolution, it is instructive to examine a simple flat ($k=0$) model of the universe, consisting of cosmic dust and the cosmological constant. The behavior of the $q$ in relation to the redshift $z$ is depicted in Figures \ref{FigS1} and \ref{FigS2}. In Figure \ref{FigS1}, we maintain a constant value of $D=2.2$ and examine the impact of $\tilde\Omega^{(m)}_0$ on the variable $q$ in terms of $z$. Our observation reveals that $q$ does not respond significantly to the current values of $\tilde\Omega^{(m)}_0$. However, in Figure \ref{FigS2}, it becomes apparent that $q$ is highly sensitive to the Lévy fractional parameter. By analyzing Figure  \ref{FigS2}, we can deduce that as $\alpha$ increases, the transition from a deceleration phase to an acceleration phase occurs at lower redshifts.

\vspace{-6pt}

\section{Fractional Cosmology Versus Data}\label{section5}

It is important to mention that the analysis in this section is limited to the background level and does not take any disturbances into account. In the following section, we compare our theoretical findings with observations. We examine four significant datasets, namely SNIa, CMB, BAO, BBN, and OHD. Some articles suggest that OHD, which is obtained based on redshift, can be compared with the SNIa dataset. If one is interested, one can refer to reference~\cite{Farooq:2013hq} and other relevant sources. 
The largest combined sample of SNIa, named the Pantheon Sample, consists of 1048 SNIa in the redshift range of $[0.01, 2.3]$, as presented by Scolnic et al.~\cite{Pan-STARRS1:2017jku}. In this analysis, we use the binned Pantheon SNIa dataset, which consists of 40 data points within the $0.014 < z < 1.61$ range. We perform an analytical marginalization of the nuisance parameter in the distance estimates $M$. In ref.~\cite{Camarena:2018nbr}, the introduction of $\chi^2_{SN}$ as
\begin{equation}
    \chi_{SN}^2(p)=S_2(p)-\frac{S_1^2(p)}{S_0}+\ln\frac{S_0}{2\pi}+\ln|2\pi S_{SN}|,
\end{equation}
in which
\begin{equation}
    \begin{split}
        S_0=V.S_{SN}^{-1}.V^T,~~~
        S_1=W.S_{SN}^{-1}.V^T,~~~
        S_2=W.S_{SN}^{-1}.W^T,
    \end{split}
\end{equation}
the covariance matrix, denoted by $S$, is calculated using the binned Pantheon sample, taking into account both statistical and systematic errors. Additionally, $V$ is a row vector consisting of unitary elements, while $W_i$ represents the difference between $\mu_{b, i}$ and $\mu(z_i)$. The distance modulus, $\mu$, is predicted theoretically using the luminosity distance $d_L$, which is given by the following equation: 
\begin{equation}
    \mu(z)=5\log_{10}\left[(1+z)\int^z_0\frac{dx}{E(x)}\right]+42.384-5\log_{10} h-19.37.
\end{equation}

Here, $E(x)$ is the dimensionless Hubble parameter, and $h$ is the Hubble constant.

The normalization constants independent of the cosmological model can be eliminated for this particular dataset. Afterwards, we incorporate the data obtained by observing acoustic signatures within the large-scale clustering of galaxies. By utilizing the BAO data, it is possible to minimize the $\chi_{BAO}^2$ that is defined as~\cite{WMAP:2012nax}
\begin{equation}
    \chi_{BAO}^2=Y^T  C_{BAO}^{-1} Y.
\end{equation}

Since the SNIa and BAO data contain valuable information regarding the universe at lower redshifts, we incorporate the CMB shift information by considering the compressed CMB likelihood. This likelihood is discussed in Table 1 of ref.~\cite{Chen:2018dbv}) and focuses on the angular scale of the sound horizon at the last scattering, denoted as $l_a$, as well as the baryon density parameter. By including this information, we aim to explore the entire history of expansion leading up to the last scattering surface. To achieve this, we refer to Equations (22)--(33) in ref.~\cite{Davari:2021mge}. Additionally, the Big Bang Nucleosynthesis (BBN) offers a data point that helps constrain mainly $\Omega^{(b)}_0$, as mentioned in the work by Serra et al.~\cite{Serra:2009yp}.

The $\chi_{BBN}^2$ is given by
\begin{equation}
    \chi_{BBN}^2=\frac{(\Omega^{(b)}_0 h^2-0.022)^2}{0.002^2}.
\end{equation}

Finally, we additionally incorporate the data obtained from the observational Hubble parameter. In this particular investigation, we utilize the 31 data points derived from the recent and precise estimates of $H(z)$ within the redshift range of $0.07 \leq z \leq 1.965$. These particular data points are independent of the Baryon Acoustic Oscillation (BAO) data points and were presented initially in~\cite{Marra:2017pst}.
In this case, we can write
\begin{equation}
    \chi_{OHD}^2(p)=\Sigma_i \frac{(H(z_i ,p)-H_i)^2}{\sigma_i^2}
\end{equation}
where $\sigma_i$ is the Gaussian error on the measured value of $H_i$.

\begin{figure}[!h]
\centering
  \includegraphics[width=8cm]{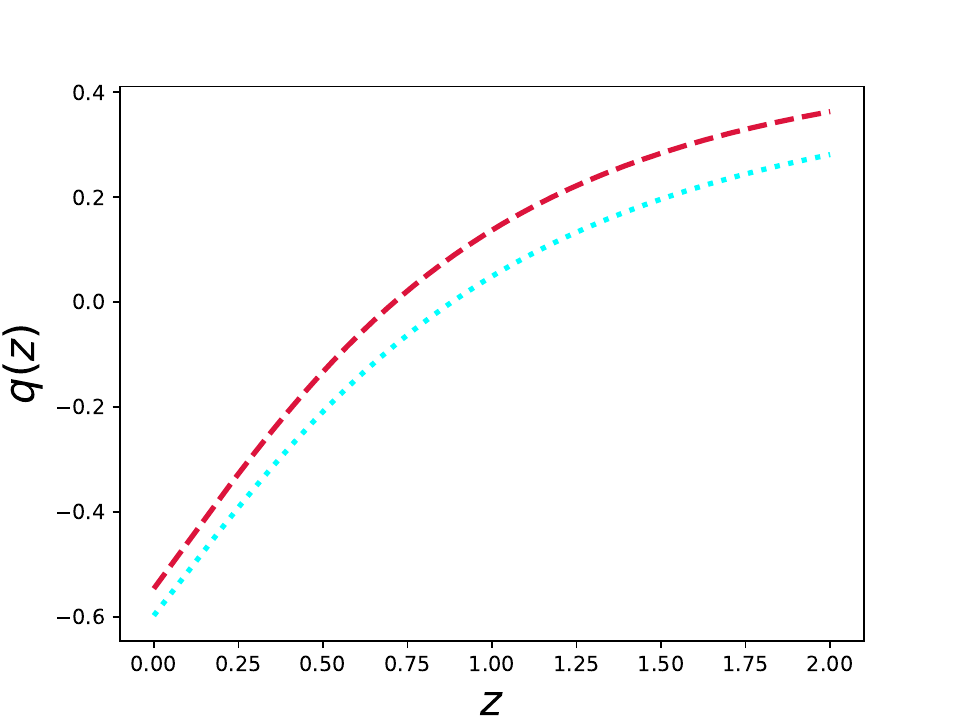}
  \caption{\small  The evolution of the deceleration parameter $q(z)$ is shown against redshift. The dotted line represents the $\Lambda$CDM model, while the dashed line represents the fractional $\Lambda$CDM model.}\label{fig4}
\end{figure}

\begin{figure}[!h]
\centering
  \includegraphics[width=11.5cm]{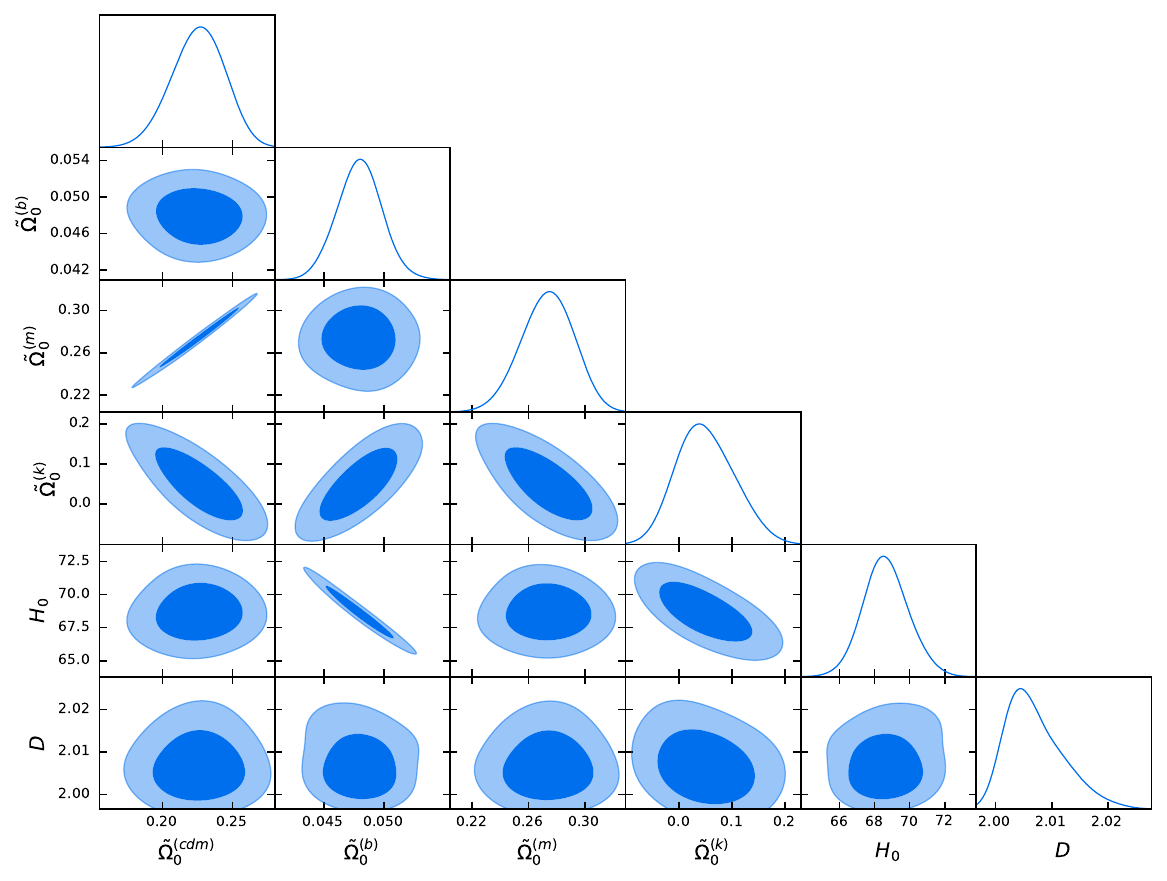}
  \caption{\small The two-dimensional contours and one-dimensional marginalized distributions represent the fractional $\Lambda$CDM model parameters with $68\%$ CL and $95\%$ CL, respectively.}\label{fig1}
\end{figure}

\begin{figure}[!h]
\centering
  \includegraphics[width=11.5cm]{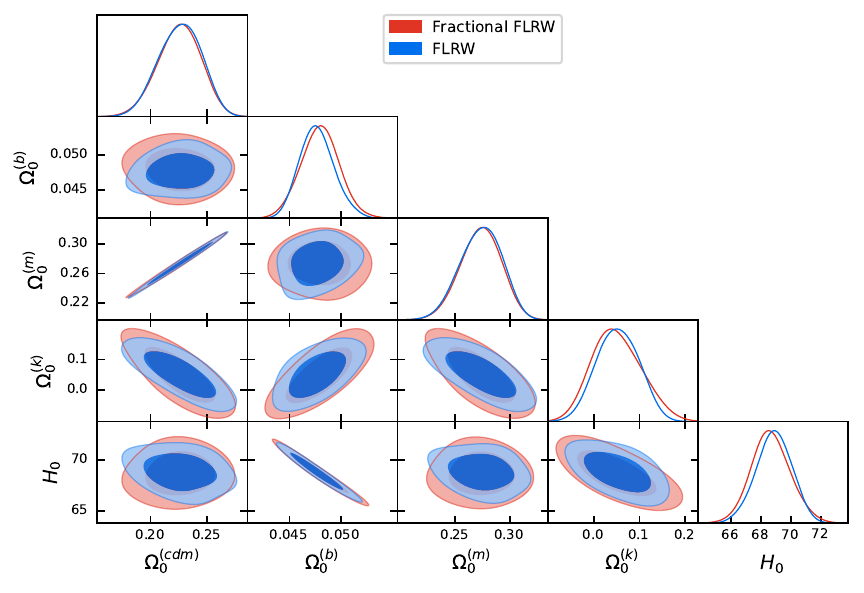}
  \caption{\small The $68\%$ and $95\%$ constraint contours on the cold dark matter density parameter $\Omega_0^{(cdm)}$, baryon density parameter $\Omega_0^{(b)}$, matter density parameter $\Omega_0^{(m)}$, curvature density parameter $\Omega_0^{(k)}$, and Hubble parameter $H_0$ when using data from the  BAO+SN+CMB+BBN+OHD.}\label{fig2}
\end{figure}

The $\Lambda$CDM model, also known as the Concordance Model, is widely accepted in cosmology as it accurately fits current observations. It has been extensively tested using various cosmological measurements. Our statistical analysis results are presented in Table \ref{Table2}
. The fractional $\Lambda$CDM model, on the other hand, introduces an additional independent parameter ($\alpha$ or $D$) that represents the attribute of nonlocality due to the fractional derivative. 
The statistical analysis of this model is presented in Table \ref{Table1}. Figure \ref{fig4} shows the evolution of the deceleration parameter $q(z)$ against redshift for the $\Lambda$CDM and fractional $\Lambda$CDM models. As we see from this figure, the deceleration parameter of the standard model of cosmology is relatively greater than the fractional deceleration parameter in all redshifts.  Figure \ref{fig1} illustrates the $1\sigma$ and $2\sigma$ confidence regions that are obtained from fitting the fractional $\Lambda$CDM model to a comprehensive set of datasets, including Baryon Acoustic Oscillations (BAO), Supernovae (SN), Cosmic Microwave Background (CMB), Big Bang Nucleosynthesis (BBN), and Observational Hubble Data (OHD). This fitting process determines the best possible agreement between the model and the observed data. In order to further evaluate the accuracy of the fractional $\Lambda$CDM model, Table \ref{Table1} was prepared to showcase the consistency between the derived values of $\tilde\Omega_m$ and $H_0$ obtained from the BAO+SN+CMB+BBN+OHD datasets and the corresponding values reported by the Planck 2018 Collaboration. By comparing the two, we can gain valuable insights into the reliability of our fits. Additionally, Figure \ref{fig2} was included to provide a visual representation of the performance of our fits as compared with the well-established $\Lambda$CDM model. It is worth mentioning that both the fractional $\Lambda$CDM model and the $\Lambda$CDM model exhibit agreement with the most up-to-date cosmological observations when it comes to their respective values of the deceleration parameter $q_0$ and the associated transition redshift $z_t=0.72$. This further solidifies the credibility of our findings.

\begin{table}[!h]
\begin{center}
\caption{The best-fit parameters with $1\sigma$ and $2\sigma$ confidence levels (CLs) for the FLRW model.}\label{Table2}
\begin{tabular}{|c|c|c|c|}
\hline
 \textbf{Parameter} &  \textbf{68\% CL }& \textbf{95\% CL} & \textbf{Best-Fit Value}\\
 \hline
{$\Omega_0^{(cdm)}   $} & $0.226^{+0.020}_{-0.018}   $ & $0.226^{+0.033}_{-0.036}   $ & $0.226^{+0.044}_{-0.048}   $\\

{$\Omega_0^{(b)}     $} & $0.0476^{+0.0015}_{-0.0018}$ & $0.0476^{+0.0034}_{-0.0032}$ & $0.0476^{+0.0045}_{-0.0037}$ \\

{$\Omega_0^{(m)}     $} & $0.274^{+0.021}_{-0.018}   $ & $0.274^{+0.034}_{-0.037}   $& $0.274^{+0.040}_{-0.050}   $\\

{$\Omega_0^{(k)}     $} & $0.052\pm 0.045            $&$0.052^{+0.089}_{-0.089}   $ & $0.05^{+0.11}_{-0.11}      $\\

{$H_0            $} & $68.9\pm 1.2               $ &  $68.9^{+2.3}_{-2.3}        $& $68.9^{+2.7}_{-3.0}        $\\
{$\chi^2_{min} $} & $68.6^{+1.3}_{-3.3}        $&$68.6^{+5.6}_{-4.2}  $ & $69^{+9}_{-5} $\\
\hline
\end{tabular}
\end{center}
\end{table}

\begin{table}[!h]
\centering
\begin{center}
\caption{The best-fit parameters with $1\sigma$ and $2\sigma$ confidence levels (CLs) for fractional FLRW model.}\label{Table1}
\begin{tabular}{|c|c|c|c|}
\hline
 \textbf{Parameter} &  \textbf{68\% CL} & \textbf{95\% CL }& \textbf{Best-Fit Value}\\
 \hline
{$\tilde\Omega_0^{(cdm)}   $} & $0.226\pm 0.018 $ & $0.226^{+0.036}_{-0.038}   $  & $0.226^{+0.046}_{-0.050}   $\\

{$\tilde\Omega_0^{(b)}     $} & $0.0480\pm 0.0018 $ & $0.0480^{+0.0036}_{-0.0036}$ & $0.0480^{+0.0052}_{-0.0045}$ \\

{$\tilde\Omega_0^{(m)}     $} & $0.274\pm 0.018 $& $0.274^{+0.032}_{-0.036}   $& $0.274^{+0.045}_{-0.049}   $\\

{$\tilde\Omega_0^{(k)}     $} & $0.049^{+0.050}_{-0.060}   $ & $0.05^{+0.11}_{-0.10}      $ & $0.05^{+0.14}_{-0.13}      $\\

{$H_0            $} & $68.6^{+1.2}_{-1.4}        $ &$68.6^{+2.6}_{-2.4}        $ & $68.6^{+3.5}_{-3.4}        $\\

{$D              $} & $2.0069^{+0.0036}_{-0.0061}$& $2.007^{+0.011}_{-0.0087}  $& $2.007^{+0.015}_{-0.010}   $\\

{$\chi^2_{min}   $} & $68.4^{+1.5}_{-3.3}        $ & $68.4^{+5.8}_{-4.4}        $& $68^{+9}_{-5}              $\\
\hline
\end{tabular}
\end{center}
\end{table}

\begin{figure}[!h]
\centering
  \includegraphics[width=8cm]{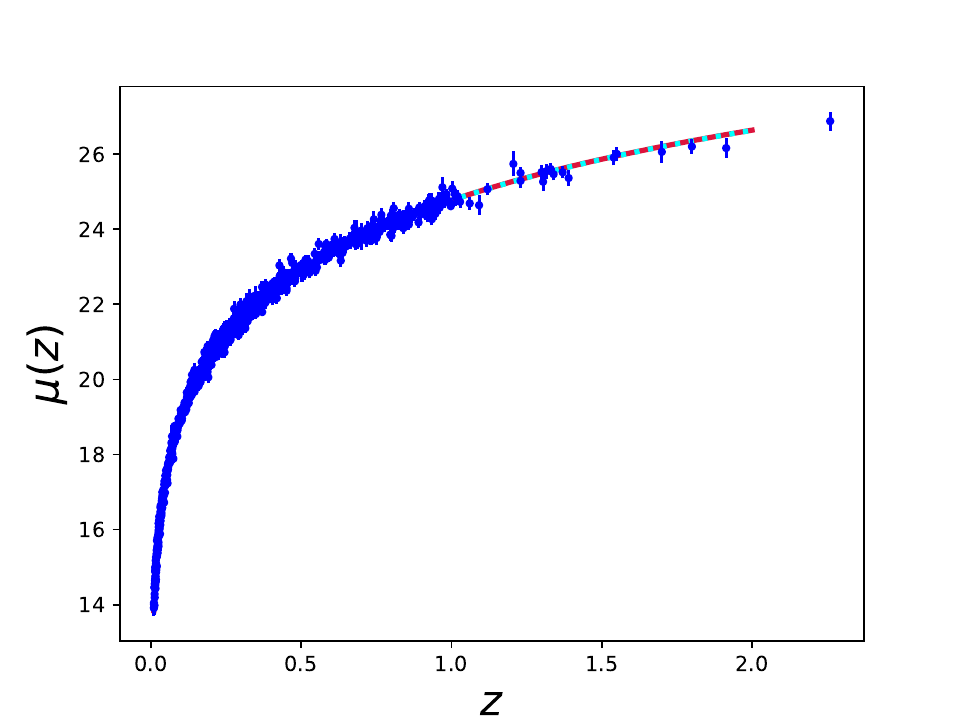}
  \caption{\small The distance modulus of supernovae (data points) is compared with the theoretically predicted distance modulus (red line) in a fractional $\Lambda$CDM model.}\label{mu}
\end{figure}

Observations of Type SNIa provide crucial data for understanding the expansion of the universe. These observations serve as primary evidence for the universe's accelerated expansion. To obtain the best possible results from teh SNIa data, we compare the observed distance modulus of SNIa detection with the theoretical value. In this regard, we utilized the Pantheon sample, which is an updated dataset of SNIa with 1048 distance moduli ($\mu$) at various redshifts within the range of $0.01 < z < 2.26$. In Figure \ref{mu}, we illustrate the Pantheon Survey as the standard Hubble diagram of SNIa with an absolute magnitude of $M_0=-19.37$. This figure also compares our model with the $\Lambda$CDM model, showing significant similarities.

\begin{figure}[!h]
\centering
\includegraphics[width=8.0cm]{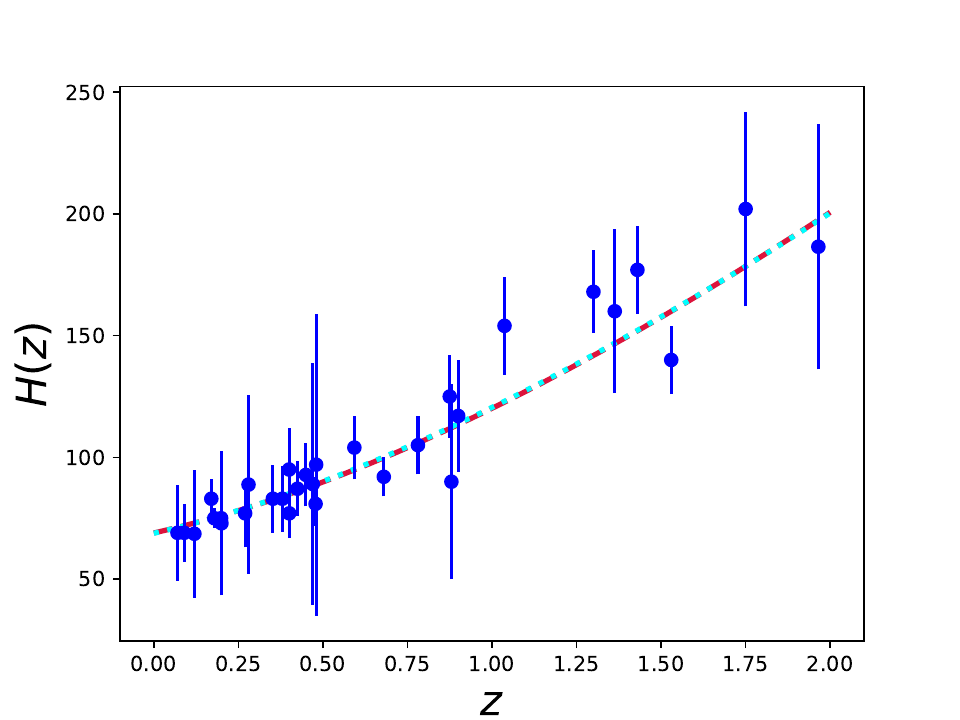}
\caption{The graph displays the evolution of the Hubble parameter, $H(z)$, in units of $\text{km}\cdot \text{sec}^{-1}\cdot \text{Mpc}^{-1}$ versus redshift $z$, with error bars. The red line represents the dynamics of the Hubble parameter obtained from Equation \eqref{Frrr}. The values of the density parameters, fractional density parameters, and Hubble constant at the present epoch are the best-fit values from Tables \ref{Table1} and \ref{Table2}, respectively.}\label{dataH}
\end{figure}

The Hubble parameter plotted against the redshift $z$ is depicted in the graph illustrated in Figure \ref{dataH}, utilizing the data extracted from Table \ref{Table2}. Upon perusing the figure, it becomes evident that our model demonstrates a remarkable degree of concurrence with the conventional model of cosmology, thereby substantiating its validity and reliability.

\section{The Synchronicity Problem }\label{section4}

Let us review the critical property of the $\Lambda$CDM model related to the synchronicity problem. The limiting case of $D=2$ in the Friedmann Equation (\ref{Frrr}) gives us the Friedmann equation in the standard ($\Lambda$CDM) model of cosmology
\begin{equation}
    \label{Frrr111}
    \left(\frac{H}{H_0}\right)^{2}=\sum_i\frac{ \Omega^{(i)}_0}{a^{3(1+\omega_i)}}+\frac{\Omega^{(k)}_0}{a^{2}},
\end{equation}
where $\Omega^{(i)}_0=\frac{8\pi G\rho^{(i)}_0}{3H_0^2}$, and $\Omega^{(k)}_0=-\frac{k}{H_0^2}$.
The solutions of the aforementioned equation for a flat universe, when radiation, cosmic dust, or the cosmological constant dominate, are given by
\begin{equation}\label{SSQQ1}
    a(t)=\begin{cases}
\left( \frac{\sqrt{H_0\Omega^{(r)}_0}}{2}\right)^\frac{1}{2}t^\frac{1}{2},~~~~\omega=\frac{1}{3},\\
\left( \frac{2\sqrt{H_0\Omega^{(m)}_0}}{3}\right)^\frac{2}{3}t^\frac{2}{3},~~~~\omega=0,\\
C\exp{\left(\sqrt{\frac{3}{\Lambda}}t \right)},~~~~\omega=-1,
    \end{cases}
\end{equation}
where, in the last equality, $C$ is a constant of integration. 
It is easy to verify that we have $Ht=1/2$ and $Ht=2/3$ in the first two cases, respectively. In the context of power-law expansion, the ratio of the universe's age to the Hubble time is a constant. This is a general property. However, when the cosmological constant dominates in late-time cosmology, we have the equation $tH = \sqrt{\Lambda/3}t$. This means that at the present epoch, concerning cosmological observations, we have $t_0\propto\sqrt{3/\Lambda}$. This equation, in fact, is an expression of the synchronicity problem of the universe’s age.

Let us analyze the effect of the Lévy fractional parameter on the synchronicity problem of the age of the universe. We obtain explicit solutions for the scale factor in a flat universe with a single-component perfect fluid. Thus, let us consider the $i$-th component is dominated in the Friedmann equation (\ref{Frrr}). Regarding Equations (\ref{Frrr}) and (\ref{decel}), the scale factor and the deceleration parameter for a perfect fluid with the equation of state $p^{(i)}=\omega_i \rho^{(i)}$ are
\begin{equation}
    \label{sol1}
    a(t)=\left(\frac{3H_0[1+\omega_i(3-D)]\tilde\Omega^{(i)}_0}{2} \right)^\frac{2}{3(1+\omega_i(3-D))}t^\frac{2}{3(1+\omega_i(3-D))},
\end{equation}
\begin{equation}\label{decelll}
    q_i=\frac{1+3\omega_i(3-D)}{2}.
\end{equation}

 This demonstrates that within the radiation-dominated universe ($\omega=1/3$), the expansion rate is $a(t)\propto t^\frac{3}{6-D}$. Furthermore, it is noteworthy that the nature of the scale factor in a matter-dominated universe ($\omega=0$) remains unaffected by the Lévy fractional parameter and is given by $a\propto t^\frac{2}{3}$. On the other hand, for a de Sitter universe ($\omega=-1$) and $D\neq2$, the above form of scale factor reduces to $a(t)\propto t^\frac{2}{3(D-2)}$. Furthermore, according to Equation (\ref{decelll}), it can be inferred that in the context of fractional cosmology, our universe has experienced a period of deceleration during the phases dominated by radiation and pressureless matter. To summarize, when considering the presence of radiation and pressureless matter, fractional cosmology is unable to elucidate the cosmic phase transition from deceleration to acceleration that occurred throughout the history of the universe unless the inclusion of dark energy (in our model, a cosmological constant) is taken into consideration. Also, the deceleration parameter (\ref{decelll}) shows that we have an accelerated universe for $D<8/3$ restricted values of $D$.

Solution (\ref{sol1}) shows that in the fractional $\Lambda$CDM model, the universe follows a power-law accelerated expansion even in recent times, unlike the standard model, where the late-time acceleration is explained by exponential de Sitter expansion. In the fractional $\Lambda$CDM model, for the $i$-th component of cosmic fluid, we have
\begin{equation}
    Ht=\frac{2}{3(1+\omega_i(3-D))}.
\end{equation}

Thus, for a de Sitter universe, we find
\begin{equation}
    Ht=\frac{2}{3(D-2)}.
\end{equation}

\begin{figure}[!h]
\centering
 \includegraphics[width=8cm]{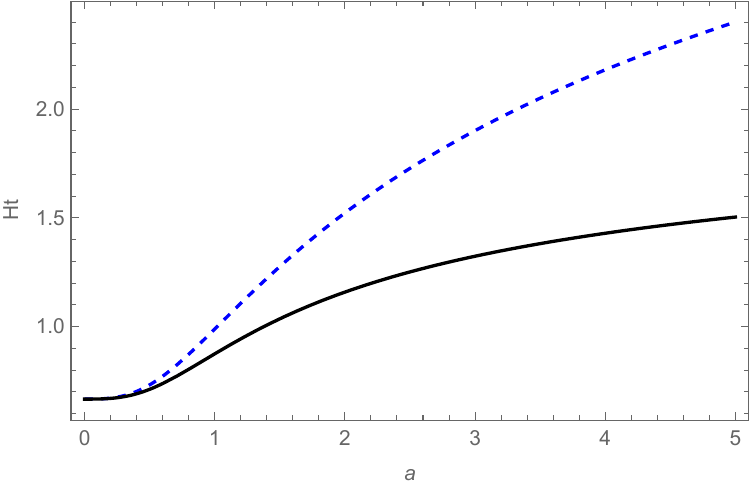}
  \caption{\small The evolution of $Ht$ with the scale factor is depicted for $D=2$ (dashed) and $D=2.3$ (line). The best-fit values for the density parameters and fractional density parameters are substituted from Tables \ref{Table1} and \ref{Table2}, respectively.}\label{HT}
\end{figure}

Unlike the $\Lambda$CDM model, the present epoch in the history of the universe is not a particular moment; instead, the age is always proportional to the Hubble time at that moment. Figure \ref{HT} displays the evolution of $Ht$ in relation to the scale factor for both the standard model of cosmology and its fractional extension. It is evident that at the current epoch, both models yield nearly identical values for $H_0t_0$. However, in the future, their behaviors will diverge. In the $\Lambda$CDM model, $Ht$ increases for $a>1$, whereas in the case of fractional $\Lambda$CDM, it remains constant, as we observe in the aforementioned single-component fluid.

The content matter of the universe has been predominantly composed of cosmic dust throughout most of its entire history. 
Regarding this fact and ignoring the radiation and curvature terms (cold matter plus the cosmological constant), one can easily show that
\begin{equation}
    Ht=f(a)\int_0^a\frac{dx}{xf(x)},~~~~~f(x)=x^\frac{3(2-D)}{2}\left(\frac{\tilde\Omega^{(m)}_0}{x^3}+1-\tilde\Omega_0^{(m)} \right)^\frac{3-D}{2}.
\end{equation}

Utilizing the above equation, we find that when $2\leq D<3$, the current age of the universe, in units of Hubble time, is restricted to
\begin{equation}\label{HT321}
  0.6667< H_0t_0\leq  0.9887.
\end{equation}

\begin{figure}[!h]
\centering
 \includegraphics[width=8cm]{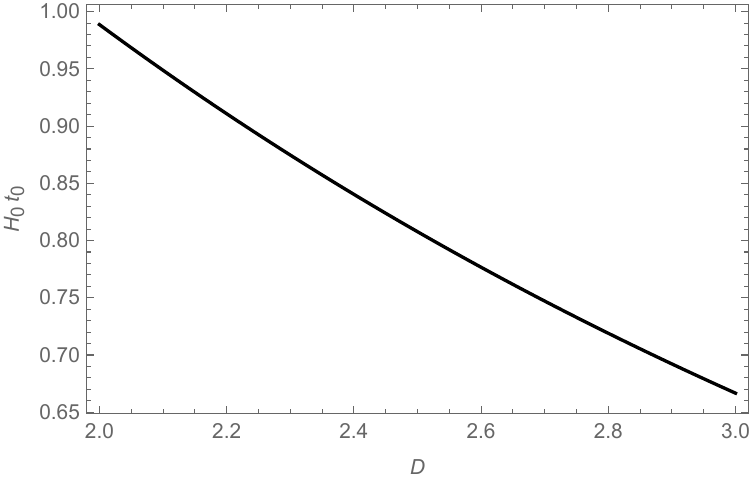}
  \caption{\small The dependence of $H_0t_0$ to the fractional parameter, $D$. The best-fit value of the fractional density parameter of dust is substituted from Table \ref{Table1}.}\label{HT2}
\end{figure}

Figure \ref{HT2} illustrates how the universe's current age is restricted for different fractional parameter values. Based on inequality (\ref{HT321}), we can observe that when the fractional parameter $D$ approaches 2, the value of $H_0t_0$ becomes almost equal to one. However, in the standard model of cosmology, where $D=2$, $Ht$ tends towards infinity in the distant future. For other fractional parameter values, $Ht$ remains finite. Based on the information presented in Table \ref{Table1}, the fractional parameter has a fitted value of $D=2.007$. For this value of $D$, we obtain
\begin{equation}
    H_0t_0=0.9858,~~~~~\lim_{a\rightarrow\infty}Ht=95.238,~~~~~t_0=13.8196~\text{Gyr}.
\end{equation}

In addition, in the upper limit of the fractional parameter, $D\rightarrow3$, we have

\begin{equation}
    \lim_{a\rightarrow\infty}Ht=0.6667.
\end{equation}

The analysis conducted shows that in the fractional $\Lambda$CDM model, the value of $Ht$ is always finite, and at the present epoch, its value is consistently close to unity. Therefore, it can be confidently concluded that the value of $Ht$ is finite in the fractional $\Lambda$CDM model.

 It is worth mentioning that fractional cosmology may have a considerable impact on an inflationary phase.
 Utilizing a scalar field with an appropriate and dominant potential compared with the universe's initial energy density is the typical method employed to achieve inflation. This inflationary approach triggers a rapid expansion of the scale factor of the universe. In the initial theories of inflation, this expansion would manifest as exponential growth, resembling de Sitter space, as the scalar field gradually moves towards its global minimum. By considering that the potential of the scalar field primarily affects inflation, we can deduce, based on the fractional deceleration parameter, that during the early universe, acceleration was experienced when the value of $D$ lay between 2 and 2.667. 
This specific range of $D$ values establishes a power-law inflationary model, and within this model, the fractional parameter $\alpha$ assumes a significant role. If we assume that the inflationary epoch took place within a time interval of $t_i=10^{-36}$ s, and $t_f = 10^{-33}$ s, we can calculate the number of e-foldings: 
\begin{equation}
    \label{inflation}
    N_e=\ln\left(\frac{a(t_f)}{a(t_i)}\right)=\frac{2}{3(D-2)}\ln\left(\frac{t_f}{t_i}\right)=\frac{2\ln(10)}{D-2}.
\end{equation}

We obtain $D\leq 2.072$ or $\alpha \geq 1.866$ if we assume a minimum of 64 e-foldings for addressing common Big Bang issues. 
This conclusion is similar to what we discovered in \ref{section5}.


\section{Conclusions}\label{conclusions}

The concept of synchronicity is perplexing because the dimensionless cosmic time, represented as $H_0t_0$, can vary significantly within the standard model of cosmology. This model explains the late-time accelerated expansion as a de Sitter universe. In this model, the current coincidence of cosmic time and  Hubble time is only valid for the present moment.
However, the age of the universe will not be as near to the Hubble time in the past or future, and $tH$ might even be infinite in the future.

In this paper, we present a comprehensive demonstration indicating that in the fractional $\Lambda$CDM model, during the late times, when the cosmological constant is dominated, the universe exhibits a remarkable and intriguing phenomenon characterized by a power-law accelerated expansion.
This stands in sharp contrast to the standard model, in which an exponential de Sitter expansion is obtained to explain the late-time acceleration.
Therefore, the age of the universe is continuously linked to the Hubble time at any given moment, creating a proportional relationship between the two phenomena. This contrasts the traditional cosmology framework, where the present epoch in the vast expanse of the universe is perceived as a specific instant in time.

Our statistical analysis  revealed that the fractional parameter proposed by L\'{e}vy, which characterizes the degree of nonlocality in a system, is determined to be precisely $\alpha=1.986$. We have also determined the age of the universe through the meticulous study of various cosmological phenomena and the use of sophisticated techniques. Our calculations show that the age of the universe is $t_0=13.8196$ billion years. Furthermore, by calculating the dimensionless age of the universe, we find that $H_0t_0=0.9858$, and $Ht=95.238$ for the distant future. These findings indicate that our model is a plausible and valid cosmological model. It successfully encompasses a crucial early stage of matter domination, essential for forming and emerging intricate structures within the universe. Furthermore, our model also incorporates a more recent acceleration phase that corresponds harmoniously with the observations made in the field of cosmology.

In conclusion, it is essential to emphasize that our fractional model, which serves as a modified gravity theory, is currently in its nascent stage and necessitates further exploration and development. By delving into numerous issues within modern cosmology, such as the Hubble tension, gravitational lensing, gravitational-wave propagation, and density perturbation generation, we not only facilitate the advancement of fractional models but also potentially uncover fascinating findings at both the quantum and classical levels. 
However, it is prudent to defer the aforementioned evaluations to future endeavors.

\vspace{6pt}

\section{ACKNOWLEDGMENTS}

We gratefully acknowledge the anonymous reviewers for their insightful comments. S.J. acknowledges financial support from the National Council for Scientific and Technological Development, Brazil--CNPq, Grant no. 308131/2022-3. S. M. M. R. acknowledges the FCT grants UID-B-MAT/00212/2020 and UID-P-MAT/00212/2020 at CMA-UBI plus the COST Action CA18108 (Quantum gravity phenomenology in the multimessenger approach).


\begin{thebibliography}{99}
\bibitem[Einstein and de~Sitter(1932)]{Einstein_1932}
Einstein, A.; de~Sitter, W.
\newblock {On the Relation between the Expansion and the Mean Density of the
  Universe}.
\newblock {\em Proc. Nat. Acad. Sci. USA} {\bf 1932}, {\em 18},~213--214.
\newblock \url{https://doi.org/10.1073/pnas.18.3.213}.

\bibitem[Chaboyer et~al.(1996)Chaboyer, Kernan, Krauss, and
  Demarque]{Chaboyer_1995}
Chaboyer, B.; Kernan, P.J.; Krauss, L.M.; Demarque, P.
\newblock {A Lower limit on the age of the universe}.
\newblock {\em Science} {\bf 1996}, {\em 271},~957--961.
\newblock {\url{https://doi.org/10.1126/science.271.5251.957}}.

\bibitem[Avelino and Kirshner(2016)]{Arturo_2016}
Avelino, A.; Kirshner, R.P.
\newblock The dimensionless age of the Universe: A riddle for our time.
\newblock {\em Astrophys. J.} {\bf 2016}, {\em 828},~35.
\newblock {\url{https://doi.org/10.3847/0004-637X/828/1/35}}.

\bibitem[Tonry et~al.(2003)]{SupernovaSearchTeam:2003cyd}
Tonry, J.L.;  et~al.
\newblock {Cosmological results from high-z supernovae}. {\em Astrophys. J.} {\bf 2003}, {\em 594},~1--24.
\newblock {\url{https://doi.org/10.1086/376865}}.

\bibitem[Casado(2020)]{Juan_2020}
Casado, J.
\newblock Linear expansion models vs. standard cosmologies: A critical and
  historical overview.
\newblock {\em Astrophys. Space Sci} {\bf 2020}, {\em 365},~16.
\newblock {\url{https://doi.org/10.1007/s10509-019-3720-z}}.

\bibitem[Tian(2022)]{Tian_2022}
Tian, S.X.
\newblock Cosmological consequences of a scalar field with oscillating equation
  of state. IV. Primordial nucleosynthesis and the deuterium problem.
\newblock {\em Phys. Rev. D} {\bf 2022}, {\em 106},~043524.
\newblock {\url{https://doi.org/10.1103/PhysRevD.106.043524}}.

\bibitem[Albuquerque et~al.(2022)Albuquerque, Frusciante, and
  Martinelli]{Albuquerque_2021}
Albuquerque, I.S.; Frusciante, N.; Martinelli, M.
\newblock {Constraining cosmological scaling solutions of a Galileon field}.
\newblock {\em Phys. Rev. D} {\bf 2022}, {\em 105},~044056.
\newblock {\url{https://doi.org/10.1103/PhysRevD.105.044056}}.

\bibitem[Kleidis and Spyrou(2016)]{Kleidis_2016}
Kleidis, K.; Spyrou, N.K.
\newblock {Dark Energy: The Shadowy Reflection of Dark Matter?} {\em Entropy} {\bf 2016}, {\em 18},~94.
\newblock {\url{https://doi.org/10.3390/e18030094}}.

\bibitem[Kleidis and Spyrou(2017)]{Kleidis_2017}
Kleidis, K.; Spyrou, N.K.
\newblock {Cosmological perturbations in the $\Lambda$CDM-like limit of a
  polytropic dark matter model}.
\newblock {\em Astron. Astrophys.} {\bf 2017}, {\em 606},~A116.
\newblock {\url{https://doi.org/10.1051/0004-6361/201630364}}.

\bibitem[Joseph and Saha(2022)]{Joseph_2022}
Joseph, A.; Saha, R.
\newblock {Forecast Analysis on Interacting Dark Energy Models from Future
  Generation PICO and DESI Missions}.
\newblock {\em Mon. Not. Roy. Astron. Soc.} {\bf 2022}, {\em 519},~1809--1822.
\newblock {\url{https://doi.org/10.1093/mnras/stac3586}}.

\bibitem[Teixeira et~al.(2022)Teixeira, Barros, Ferreira, and
  Frusciante]{Teixeira_2022}
Teixeira, E.M.; Barros, B.J.; Ferreira, V.M.C.; Frusciante, N.
\newblock {Dissecting kinetically coupled quintessence: Phenomenology and
  observational tests}.
\newblock {\em JCAP} {\bf 2022}, {\em 11},~059.
\newblock {\url{https://doi.org/10.1088/1475-7516/2022/11/059}}.

\bibitem[Liu et~al.(2022)Liu, Liao, Liu, Zhang, An, and Fan]{Liu_2022}
Liu, Y.; Liao, S.; Liu, X.; Zhang, J.; An, R.; Fan, Z.
\newblock {Dark matter haloes in interacting dark energy models: Formation
  history, density profile, spin, and shape}.
\newblock {\em Mon. Not. Roy. Astron. Soc.} {\bf 2022}, {\em 511},~3076--3088.
\newblock {\url{https://doi.org/10.1093/mnras/stac229}}.

\bibitem[Nayak(2020)]{Nayak_2019}
Nayak, B.
\newblock {Interacting Holographic Dark Energy, the Present Accelerated
  Expansion and Black Holes}.
\newblock {\em Grav. Cosmol.} {\bf 2020}, {\em 26},~273--280.
\newblock {\url{https://doi.org/10.1134/S020228932003010X}}.

\bibitem[Escamilla-Rivera and Gamboa(2022)]{Rivera_2022}
Escamilla-Rivera, C.; Gamboa, A.
\newblock {A new parameterized interacting holographic dark energy}.
\newblock {\em Eur. Phys. J. Plus} {\bf 2022}, {\em 137},~254.
\newblock {\url{https://doi.org/10.1140/epjp/s13360-022-02490-4}}.

\bibitem[Landim(2022)]{Landim_2022}
Landim, R.G.
\newblock {Note on interacting holographic dark energy with a Hubble-scale
  cutoff}.
\newblock {\em Phys. Rev. D} {\bf 2022}, {\em 106},~043527.
\newblock {\url{https://doi.org/10.1103/PhysRevD.106.043527}}.

\bibitem[{Shimon}(2024)]{Shimon_2022}
{Shimon}, M.
\newblock {Elucidation of 'Cosmic Coincidence'} \emph{New Astron.} {\bf 2024}.
\newblock {\em 106},~102126.
\newblock {\url{https://doi.org/10.1016/j.newast.2023.102126}}.

\bibitem[Anari and Sadjadi(2022)]{Anari_2022}
Anari, V.; Sadjadi, H.M.
\newblock {Constant-roll, cosmic acceleration, and massive neutrinos}.
\newblock {\em JCAP} {\bf 2022}, {\em 7},~043.
\newblock {\url{https://doi.org/10.1088/1475-7516/2022/07/043}}.

\bibitem[Jalalzadeh et~al.(2023)Jalalzadeh, Mohammadi, and
  Demir]{Jalalzadeh_2022b}
Jalalzadeh, S.; Mohammadi, A.; Demir, D.
\newblock {A quantum cosmology approach to cosmic coincidence and inflation}.
\newblock {\em Phys. Dark Univ.} {\bf 2023}, {\em 40},~101227.
\newblock {\url{https://doi.org/10.1016/j.dark.2023.101227}}.

\bibitem[Kolb(1989)]{Kolb_1989}
Kolb, E.W.
\newblock A coasting cosmology.
\newblock {\em Astrophys. J.} {\bf 1989}, {\em 344}, 543--550.
\newblock {\url{https://doi.org/10.1086/167825}}.

\bibitem[Allen(1999)]{Allen_1999}
Allen, R.E.
\newblock {Four testable predictions of instanton cosmology}.
\newblock {\em AIP Conf. Proc.} {\bf 1999}, {\em 478},~204--207.
\newblock {\url{https://doi.org/10.1063/1.59392}}.

\bibitem[Pimentel and D{\'i}az-Rivera(1998)]{Pimentel_1999}
Pimentel, L.O.; D{\'i}az-Rivera, L.M.
\newblock Coasting Cosmologies with Time Dependent Cosmological Constant.
\newblock {\em Int. J. Mod. Phys. A} {\bf 1998}, {\em 14},~1523--1529.
\newblock {\url{https://doi.org/10.1142/S0217751X99000762}}.

\bibitem[Melia and Shevchuk(2012)]{Melia_2012}
Melia, F.; Shevchuk, A.
\newblock The $R_h=ct$ universe.
\newblock {\em Mon. Not. Roy. Astron. Soc.} {\bf 2012}, {\em 419},~2579--2586.
\newblock {\url{https://doi.org/10.1111/j.1365-2966.2011.19906.x}}.

\bibitem[Lewis(2013)]{Lewis_2013}
Lewis, G.F.
\newblock Matter matters: Unphysical properties of the $R_h=ct$ universe.
\newblock {\em Mon. Not. Roy. Astron. Soc.} {\bf 2013}, {\em 432},~2324--2330.
\newblock {\url{https://doi.org/10.1093/mnras/stt592}}.

\bibitem[Moniz and Jalalzadeh(2020)]{Moniz_2020}
Moniz, P.V.; Jalalzadeh, S.
\newblock {From Fractional Quantum Mechanics to Quantum Cosmology: An
  Overture}.
\newblock {\em Mathematics} {\bf 2020}, {\em 8},~313.
\newblock {\url{https://doi.org/10.3390/math8030313}}.

\bibitem[Rasouli et~al.(2021)Rasouli, Jalalzadeh, and Moniz]{Moniz_2021}
Rasouli, S.; Jalalzadeh, S.; Moniz, P.
\newblock Broadening quantum cosmology with a fractional whirl.
\newblock {\em Mod. Phys. Lett. A} {\bf 2021}, {\em 36},~2140005.
\newblock {\url{https://doi.org/10.1142/S0217732321400058}}.

\bibitem[Jalalzadeh et~al.(2021)Jalalzadeh, da~Silva, and
  Moniz]{Jalalzadeh_2021}
Jalalzadeh, S.; da~Silva, F.R.; Moniz, P.
\newblock Prospecting black hole thermodynamics with fractional quantum
  mechanics.
\newblock {\em Eur. Phys. J. C} {\bf 2021}, {\em 81},~632.
\newblock {\url{https://doi.org/10.1140/epjc/s10052-021-09438-5}}.

\bibitem[Jalalzadeh et~al.(2022)Jalalzadeh, Costa, and Moniz]{Jalalzadeh_2022}
Jalalzadeh, S.; Costa, E.O.; Moniz, P.
\newblock de Sitter fractional quantum cosmology.
\newblock {\em Phys. Rev. D} {\bf 2022}, {\em 105},~L121901.
\newblock {\url{https://doi.org/10.1103/PhysRevD.105.L121901}}.

\bibitem[Rasouli et~al.(2022)Rasouli, Costa, Moniz, and
  Jalalzadeh]{Rasouli:2022bug}
Rasouli, S.M.M.; Costa, E.W.O.; Moniz, P.V.; Jalalzadeh, S.
\newblock {Inflation and fractional quantum cosmology}.
\newblock {\em Fractal Fract.} {\bf 2022}, {\em 6},~655.
\newblock {\url{https://doi.org/10.3390/fractalfract6110655}}.

\bibitem[Jalalzadeh and Vargas~Moniz(2022)]{Jalalzadeh_2022c}
Jalalzadeh, S.; Vargas~Moniz, P.
\newblock {\em {Challenging Routes in Quantum Cosmology}}; World Scientific: {Singapore}
,
  2022.
\newblock {\url{https://doi.org/10.1142/8540}}.

\bibitem[Laskin(2000)]{Laskin_2000}
Laskin, N.
\newblock Fractional Quantum Mechanics.
\newblock {\em Phys. Rev. E} {\bf 2000}, {\em 62},~3135--3145.
\newblock {\url{https://doi.org/10.1103/PhysRevE.62.3135}}.

\bibitem[Laskin(2002)]{Laskin_2002}
Laskin, N.
\newblock {Fractional Schrodinger equation}.
\newblock {\em Phys. Rev. E} {\bf 2002}, {\em 66},~056108.
\newblock {\url{https://doi.org/10.1103/PhysRevE.66.056108}}.

\bibitem[Laskin(2010)]{Laskin_2010}
Laskin, N.
\newblock {Principles of Fractional Quantum Mechanics}. \emph{arXiv} {\bf 2010}, arXiv:1009.5533.

\bibitem[Calcagni(2017)]{Calcagni_2017}
Calcagni, G.
\newblock {\em {Classical and Quantum Cosmology}}; Graduate Texts in Physics;
  Springer: {Berlin/Heidelberg, Germany}, 2017.
\newblock {\url{https://doi.org/10.1007/978-3-319-41127-9}}.

\bibitem[Rasouli et~al.(2011)Rasouli, Farhoudi, and Khosravi]{Rasouli:2011zkm}
Rasouli, S.M.M.; Farhoudi, M.; Khosravi, N.
\newblock {Horizon Problem Remediation via Deformed Phase Space}.
\newblock {\em Gen. Rel. Grav.} {\bf 2011}, {\em 43},~2895--2910.
\newblock {\url{https://doi.org/10.1007/s10714-011-1208-4}}.

\bibitem[Rasouli et~al.(2014)Rasouli, Ziaie, Marto, and Moniz]{Rasouli:2013sda}
Rasouli, S.M.M.; Ziaie, A.H.; Marto, J.; Moniz, P.V.
\newblock {Gravitational Collapse of a Homogeneous Scalar Field in Deformed
  Phase Space}.
\newblock {\em Phys. Rev. D} {\bf 2014}, {\em 89},~044028.
\newblock {\url{https://doi.org/10.1103/PhysRevD.89.044028}}.

\bibitem[Jalalzadeh et~al.(2014)Jalalzadeh, Rasouli, and
  Moniz]{Jalalzadeh:2014jea}
Jalalzadeh, S.; Rasouli, S.M.M.; Moniz, P.V.
\newblock {Quantum cosmology, minimal length and holography}.
\newblock {\em Phys. Rev. D} {\bf 2014}, {\em 90},~023541.
\newblock {\url{https://doi.org/10.1103/PhysRevD.90.023541}}.

\bibitem[Rasouli and Vargas~Moniz(2014)]{Rasouli:2014dba}
Rasouli, S.M.M.; Vargas~Moniz, P.
\newblock {Noncommutative minisuperspace, gravity-driven acceleration, and
  kinetic inflation}.
\newblock {\em Phys. Rev. D} {\bf 2014}, {\em 90},~083533.
\newblock {\url{https://doi.org/10.1103/PhysRevD.90.083533}}.

\bibitem[Rasouli et~al.(2016)Rasouli, Ziaie, Jalalzadeh, and
  Moniz]{Rasouli:2016xdo}
Rasouli, S.M.M.; Ziaie, A.H.; Jalalzadeh, S.; Moniz, P.V.
\newblock {Non-singular Brans\textendash{}Dicke collapse in deformed phase
  space}.
\newblock {\em Annals Phys.} {\bf 2016}, {\em 375},~154--178.
\newblock {\url{https://doi.org/10.1016/j.aop.2016.09.007}}.

\bibitem[Rasouli and Vargas~Moniz(2016)]{Rasouli:2016syh}
Rasouli, S.M.M.; Vargas~Moniz, P.
\newblock {Gravity-Driven Acceleration and Kinetic Inflation in Noncommutative
  Brans-Dicke Setting}.
\newblock {\em Odessa Astron. Pub.} {\bf 2016}, {\em 29},~19.
\newblock {\url{https://doi.org/10.18524/1810-4215.2016.29.84956}}.

\bibitem[Rasouli et~al.(2018)Rasouli, Saba, Farhoudi, Marto, and
  Moniz]{Rasouli:2018nwi}
Rasouli, S.M.M.; Saba, N.; Farhoudi, M.; Marto, J.a.; Moniz, P.V.
\newblock {Inflationary Universe in Deformed Phase Space Scenario}.
\newblock {\em Annals Phys.} {\bf 2018}, {\em 393},~288--307.
\newblock {\url{https://doi.org/10.1016/j.aop.2018.04.014}}.

\bibitem[Rasouli et~al.(2019)Rasouli, Marto, and Vargas~Moniz]{Rasouli:2018lny}
Rasouli, S.M.M.; Marto, J.a.; Vargas~Moniz, P.
\newblock {Kinetic inflation in deformed phase space Brans\textendash{}Dicke
  cosmology}.
\newblock {\em Phys. Dark Univ.} {\bf 2019}, {\em 24},~100269.
\newblock {\url{https://doi.org/10.1016/j.dark.2019.100269}}.

\bibitem[Rasouli(2022)]{Rasouli:2022hnp}
Rasouli, S.M.M.
\newblock {Noncommutativity, S\'aez\textendash{}Ballester Theory and Kinetic
  Inflation}.
\newblock {\em Universe} {\bf 2022}, {\em 8},~165.
\newblock {\url{https://doi.org/10.3390/universe8030165}}.

\bibitem[Shchigolev(2011)]{Shchigolev_2011}
Shchigolev, V.
\newblock Cosmological models with fractional derivatives and fractional action
  functional.
\newblock {\em Commun. Theor. Phys.} {\bf 2011}, {\em 56},~389.
\newblock {\url{https://doi.org/10.1088/0253-6102/56/2/34}}.

\bibitem[Varieschi(2021)]{Gabriele_2021}
Varieschi, G.U.
\newblock Relativistic Fractional-Dimension Gravity.
\newblock {\em Universe} {\bf 2021}, {\em 7},~387.
\newblock {\url{https://doi.org/10.3390/universe7100387}}.

\bibitem[Garc\'\i{}a-Aspeitia et~al.(2022)Garc\'\i{}a-Aspeitia,
  Fernandez-Anaya, Hern\'andez-Almada, Leon, and Maga\~na]{Garcia_2022}
Garc\'\i{}a-Aspeitia, M.A.; Fernandez-Anaya, G.; Hern\'andez-Almada, A.; Leon,
  G.; Maga\~na, J.
\newblock {Cosmology under the fractional calculus approach}.
\newblock {\em Mon. Not. Roy. Astron. Soc.} {\bf 2022}, {\em 517},~4813--4826.
\newblock {\url{https://doi.org/10.1093/mnras/stac3006}}.

\bibitem[Leon et~al.(2023)Leon, Garc\'\i{}a-Aspeitia, Fernandez-Anaya,
  Hern\'andez-Almada, Maga\~na, and Gonz\'alez]{Leon_2023}
Leon, G.; Garc\'\i{}a-Aspeitia, M.A.; Fernandez-Anaya, G.; Hern\'andez-Almada,
  A.; Maga\~na, J.; Gonz\'alez, E.
\newblock {Cosmology under the fractional calculus approach: A possible $H_0$
  tension resolution?} \emph{arXiv} {\bf 2023}, arXiv:2304.14465.

\bibitem[Shchigolev(2013{\natexlab{a}})]{Shchigolev:2012rp}
Shchigolev, V.K.
\newblock {Cosmic Evolution in Fractional Action Cosmology}.
\newblock {\em Discontinuitynlinearity Complex.} {\bf 2013}, {\em
  2},~115--123.
\newblock {\url{https://doi.org/10.5890/DNC.2013.04.002}}.

\bibitem[Shchigolev(2013{\natexlab{b}})]{Shchigolev:2013jq}
Shchigolev, V.K.
\newblock {Fractional Einstein-Hilbert Action Cosmology}.
\newblock {\em Mod. Phys. Lett. A} {\bf 2013}, {\em 28},~1350056.
\newblock {\url{https://doi.org/10.1142/S0217732313500569}}.

\bibitem[Calcagni(2013)]{Calcagni:2013yqa}
Calcagni, G.
\newblock {Multi-scale gravity and cosmology}.
\newblock {\em JCAP} {\bf 2013}, {\em 12},~041.
\newblock {\url{https://doi.org/10.1088/1475-7516/2013/12/041}}.

\bibitem[Shchigolev(2016)]{Shchigolev:2015rei}
{Shchigolev, V.K.
\newblock {Testing Fractional Action Cosmology}.
\newblock {\em Eur. Phys. J. Plus} {\bf 2016}, {\em 131},~256.
\newblock {\url{https://doi.org/10.1140/epjp/i2016-16256-6}}.}

\bibitem[Calcagni(2017)]{Calcagni:2016azd}
Calcagni, G.
\newblock {Multifractional theories: An unconventional review}.
\newblock {\em JHEP} {\bf 2017}, {\em 3},~138. Erratum in \textbf{2017}, \emph{6}, 20.
  {\url{https://doi.org/10.1007/JHEP03(2017)138}}.

\bibitem[Shchigolev(2021)]{Shchigolev:2021lbm}
Shchigolev, V.K.
\newblock {Fractional-order derivatives in cosmological models of accelerated
  expansion}.
\newblock {\em Mod. Phys. Lett. A} {\bf 2021}, {\em 36},~2130014.
\newblock {\url{https://doi.org/10.1142/S0217732321300147}}.

\bibitem[Calcagni and De~Felice(2020)]{Calcagni:2020ads}
Calcagni, G.; De~Felice, A.
\newblock {Dark energy in multifractional spacetimes}.
\newblock {\em Phys. Rev. D} {\bf 2020}, {\em 102},~103529.
\newblock {\url{https://doi.org/10.1103/PhysRevD.102.103529}}.

\bibitem[Calcagni(2021{\natexlab{a}})]{Calcagni:2021ipd}
Calcagni, G.
\newblock {Multifractional theories: An updated review}.
\newblock {\em Mod. Phys. Lett. A} {\bf 2021}, {\em 36},~2140006.
\newblock {\url{https://doi.org/10.1142/S021773232140006X}}.

\bibitem[Calcagni(2021{\natexlab{b}})]{Calcagni:2021aap}
Calcagni, G.
\newblock {Classical and quantum gravity with fractional operators}.
\newblock {\em Class. Quant. Grav.} {\bf 2021}, {\em 38},~165005. Erratum in \textbf{2021}, \emph{38}, 169601.
  {\url{https://doi.org/10.1088/1361-6382/ac1bea}}.

\bibitem[Gonz\'alez et~al.(2023)Gonz\'alez, Leon, and
  Fernandez-Anaya]{Gonzalez:2023who}
Gonz\'alez, E.; Leon, G.; Fernandez-Anaya, G.
\newblock {Exact solutions and cosmological constraints in fractional
  cosmology}.
\newblock {\em Fractal Fract.} {\bf 2023}, {\em 7},~368.
\newblock {\url{https://doi.org/10.3390/fractalfract7050368}}.

\bibitem[Socorro and Rosales(2023)]{Socorro:2023ztq}
Socorro, J.; Rosales, J.J.
\newblock {Quantum Fractionary Cosmology: K-Essence Theory}.
\newblock {\em Universe} {\bf 2023}, {\em 9},~185.
\newblock {\url{https://doi.org/10.3390/universe9040185}}.

\bibitem[Calcagni and Kuroyanagi(2021)]{Calcagni:2020tvw}
Calcagni, G.; Kuroyanagi, S.
\newblock {Stochastic gravitational-wave background in quantum gravity}. {\em JCAP} {\bf 2021}, {\em 3},~019.
\newblock {\url{https://doi.org/10.1088/1475-7516/2021/03/019}}.

\bibitem[Calcagni et~al.(2019)Calcagni, Kuroyanagi, Marsat, Sakellariadou,
  Tamanini, and Tasinato]{Calcagni:2019ngc}
Calcagni, G.; Kuroyanagi, S.; Marsat, S.; Sakellariadou, M.; Tamanini, N.;
  Tasinato, G.
\newblock {Quantum gravity and gravitational-wave astronomy}.
\newblock {\em JCAP} {\bf 2019}, {\em 10},~012.
\newblock {\url{https://doi.org/10.1088/1475-7516/2019/10/012}}.

\bibitem[Calcagni(2017)]{Calcagni:2017via}
Calcagni, G.
\newblock {Complex dimensions and their observability}.
\newblock {\em Phys. Rev. D} {\bf 2017}, {\em 96},~046001.
\newblock {\url{https://doi.org/10.1103/PhysRevD.96.046001}}.

\bibitem[Calcagni et~al.(2016)Calcagni, Kuroyanagi, and
  Tsujikawa]{Calcagni:2016ofu}
Calcagni, G.; Kuroyanagi, S.; Tsujikawa, S.
\newblock {Cosmic microwave background and inflation in multi-fractional
  spacetimes}.
\newblock {\em JCAP} {\bf 2016}, {\em 8},~039.
\newblock {\url{https://doi.org/10.1088/1475-7516/2016/08/039}}.

\bibitem[El-Nabulsi(2012)]{El-Nabulsi:2012wpc}
El-Nabulsi, R.A.
\newblock {Gravitons in fractional action cosmology}.
\newblock {\em Int. J. Theor. Phys.} {\bf 2012}, {\em 51},~3978--3992.
\newblock {\url{https://doi.org/10.1007/s10773-012-1290-8}}.

\bibitem[El-Nabulsi(2016)]{El-Nabulsi:2015szp}
El-Nabulsi, R.A.
\newblock {A Cosmology Governed by a Fractional Differential Equation and the
  Generalized Kilbas-Saigo-Mittag-Leffler Function}.
\newblock {\em Int. J. Theor. Phys.} {\bf 2016}, {\em 55},~625--635.
\newblock {\url{https://doi.org/10.1007/s10773-015-2700-5}}.

\bibitem[Jamil et~al.(2012)Jamil, Momeni, and Rashid]{Jamil:2011uj}
Jamil, M.; Momeni, D.; Rashid, M.A.
\newblock {Fractional Action Cosmology with Power Law Weight Function}.
\newblock {\em J. Phys. Conf. Ser.} {\bf 2012}, {\em 354},~012008.
\newblock {\url{https://doi.org/10.1088/1742-6596/354/1/012008}}.

\bibitem[El-Nabulsi(2013{\natexlab{a}})]{El-Nabulsi:2013mma}
El-Nabulsi, R.A.
\newblock {Nonstandard fractional exponential Lagrangians, fractional geodesic
  equation, complex general relativity, and discrete gravity}.
\newblock {\em Can. J. Phys.} {\bf 2013}, {\em 91},~618--622.
\newblock {\url{https://doi.org/10.1139/cjp-2013-0145}}.

\bibitem[El-Nabulsi(2013{\natexlab{b}})]{El-Nabulsi:2013mwa}
El-Nabulsi, A.R.
\newblock {Non-minimal coupling in fractional action cosmology}.
\newblock {\em Indian J. Phys.} {\bf 2013}, {\em 87},~835--840.
\newblock {\url{https://doi.org/10.1007/s12648-013-0295-3}}.

\bibitem[Rami(2015)]{Rami:2015kha}
Rami, E.N.A.
\newblock {Fractional action oscillating phantom cosmology with conformal
  coupling}.
\newblock {\em Eur. Phys. J. Plus} {\bf 2015}, {\em 130},~102.
\newblock {\url{https://doi.org/10.1140/epjp/i2015-15102-9}}.

\bibitem[El-Nabulsi(2016)]{El-Nabulsi:2016dsj}
El-Nabulsi, R.A.
\newblock {Implications of the Ornstein-Uhlenbeck-like fractional differential
  equation in cosmology}.
\newblock {\em Rev. Mex. Fis.} {\bf 2016}, {\em 62},~240.

\bibitem[El-Nabulsi(2017{\natexlab{a}})]{El-Nabulsi:2017vmp}
El-Nabulsi, R.A.
\newblock {Fractional Action Cosmology with Variable Order Parameter}.
\newblock {\em Int. J. Theor. Phys.} {\bf 2017}, {\em 56},~1159--1182.
\newblock {\url{https://doi.org/10.1007/s10773-016-3260-z}}.

\bibitem[El-Nabulsi(2017{\natexlab{b}})]{El-Nabulsi:2017jss}
{El-Nabulsi, R.A.
\newblock {Wormholes in fractional action cosmology}.
\newblock {\em Can. J. Phys.} {\bf 2017}, {\em 95},~605--609.
\newblock {\url{https://doi.org/10.1139/cjp-2017-0109}}.}

\bibitem[El-Nabulsi(2017{\natexlab{c}})]{El-Nabulsi:2017rdu}
El-Nabulsi, R.A.
\newblock {New Metrics from a Fractional Gravitational Field}.
\newblock {\em Commun. Theor. Phys.} {\bf 2017}, {\em 68},~309.
\newblock {\url{https://doi.org/10.1088/0253-6102/68/3/309}}.

\bibitem[Debnath et~al.(2012)Debnath, Jamil, and Chattopadhyay]{Debnath2012}
Debnath, U.; Jamil, M.; Chattopadhyay, S.
\newblock {Fractional Action Cosmology: Emergent, Logamediate, Intermediate,
  Power Law Scenarios of the Universe and Generalized Second Law of
  Thermodynamics}.
\newblock {\em Int. J. Theor. Phys} {\bf 2012}, {\em 51},~812--837.
\newblock {\url{https://doi.org/10.1007/s10773-011-0961-1}}.

\bibitem[Debnath et~al.(2013)Debnath, Chattopadhyay, and Jamil]{Debnath2013}
Debnath, U.; Chattopadhyay, S.; Jamil, M.
\newblock {Fractional action cosmology: Some dark energy models in emergent,
  logamediate, and intermediate scenarios of the universe}.
\newblock {\em Int. J. Theor. Phys} {\bf 2013}, {\em 7},~25.
\newblock {\url{https://doi.org/10.1186/2251-7235-7-25}}.

\bibitem[Calcagni(2010{\natexlab{a}})]{Calcagni:2010bj}
Calcagni, G.
\newblock {Quantum field theory, gravity and cosmology in a fractal universe}.
\newblock {\em JHEP} {\bf 2010}, {\em 3},~120.
\newblock {\url{https://doi.org/10.1007/JHEP03(2010)120}}.

\bibitem[Calcagni(2010{\natexlab{b}})]{Calcagni:2009kc}
{Calcagni, G.
\newblock {Fractal universe and quantum gravity}.
\newblock {\em Phys. Rev. Lett.} {\bf 2010}, {\em 104},~251301.
\newblock {\url{https://doi.org/10.1103/PhysRevLett.104.251301}}.}

\bibitem[Junior et~al.(2023)Junior, Costa, and Jalalzadeh]{Junior:2023fwb}
Junior, P.F.d.S.; Costa, E.W.d.O.; Jalalzadeh, S.
\newblock {Emergence of fractal cosmic space from fractional quantum gravity}.
\newblock {\em Eur. Phys. J. Plus} {\bf 2023}, {\em 138},~862.
\newblock {\url{https://doi.org/10.1140/epjp/s13360-023-04506-z}}.

\bibitem[Barrientos et~al.(2021)Barrientos, Mendoza, and
  Padilla]{Barrientos:2020kfp}
Barrientos, E.; Mendoza, S.; Padilla, P.
\newblock {Extending Friedmann equations using fractional derivatives using a
  Last Step Modification technique: The case of a matter dominated accelerated
  expanding Universe}.
\newblock {\em Symmetry} {\bf 2021}, {\em 13},~174.
\newblock {\url{https://doi.org/10.3390/sym13020174}}.

\bibitem[Landim(2021)]{Landim:2021www}
Landim, R.G.
\newblock {Fractional dark energy}.
\newblock {\em Phys. Rev. D} {\bf 2021}, {\em 103},~083511.
\newblock {\url{https://doi.org/10.1103/PhysRevD.103.083511}}.

\bibitem[Calcagni(2021)]{Calcagni_2021}
Calcagni, G.
\newblock {Quantum scalar field theories with fractional operators}.
\newblock {\em Class. Quant. Grav.} {\bf 2021}, {\em 38},~165006.
\newblock {\url{https://doi.org/10.1088/1361-6382/ac103c}}.

\bibitem[Landim(2021)]{Landim:2021ial}
Landim, R.G.
\newblock {Fractional dark energy: Phantom behavior and negative absolute
  temperature}.
\newblock {\em Phys. Rev. D} {\bf 2021}, {\em 104},~103508.
\newblock {\url{https://doi.org/10.1103/PhysRevD.104.103508}}.

\bibitem[Giusti(2020)]{Giusti:2020rul}
{Giusti, A.
\newblock {MOND-like Fractional Laplacian Theory}.
\newblock {\em Phys. Rev. D} {\bf 2020}, {\em 101},~124029.
\newblock {\url{https://doi.org/10.1103/PhysRevD.101.124029}}.}

\bibitem[Torres et~al.(2020)Torres, Fabris, Piattella, and
  Batista]{Torres:2020xkw}
Torres, I.; Fabris, J.C.; Piattella, O.F.; Batista, A.B.
\newblock {Quantum Cosmology of Fab Four John Theory with Conformable
  Fractional Derivative}.
\newblock {\em Universe} {\bf 2020}, {\em 6},~50.
\newblock {\url{https://doi.org/10.3390/universe6040050}}.

\bibitem[Kilbas et~al.(2006)Kilbas, Srivastava, and Trujillo]{book1:2006}
Kilbas, A.; Srivastava, H.; Trujillo, J.
\newblock Theory and applications Of Fractional Differential Equations.
\newblock {\em North-Holland Math. Stud.} {\bf 2006}, {\em 204},7--10.
\newblock {\url{https://doi.org/10.1016/S0304-0208(06)80001-0}}.

\bibitem[Podlubny(1998)]{book2:1999}
Podlubny, I.
\newblock {\em Fractional Differential Equations}; Elsevier: {Amsterdam, The Netherlands}, 1998; Volume 198.

\bibitem[Laskin(2000)]{Laskin:1999tf}
Laskin, N.
\newblock {Fractional quantum mechanics and Levy paths integrals}.
\newblock {\em Phys. Lett. A} {\bf 2000}, {\em 268},~298--305.
\newblock {\url{https://doi.org/10.1016/S0375-9601(00)00201-2}}.

\bibitem[Pozrikidis(2018)]{Pozrikidis_2018}
Pozrikidis, C.
\newblock {\em The Fractional Laplacian}; {CRC Press: Boca Raton, FL, USA}, 2018.
\newblock {\url{https://doi.org/10.1201/9781315367675}}.

\bibitem[Laskin(2018)]{doi:10.1142/10541}
Laskin, N.
\newblock {\em Fractional Quantum Mechanics}; World Scientific: {Singapore},  2018.
\newblock {\url{https://doi.org/10.1142/10541}}.

\bibitem[Riesz(1949)]{Rie}
Riesz, M.
\newblock {L'intégrale de Riemann-Liouville et le problème de Cauchy}.
\newblock {\em Acta Mathematica} {\bf 1949}, {\em 81},~1 -- 222.

\bibitem[Tarasov(2018)]{Tarasov:2018zjg}
Tarasov, V.E.
\newblock {Fractional Derivative Regularization in QFT}.
\newblock {\em Adv. High Energy Phys.} {\bf 2018}, {\em 2018},~7612490.
\newblock {\url{https://doi.org/10.1155/2018/7612490}}.

\bibitem[Batalin and Vilkovisky(1977)]{Batalin:1977pb}
Batalin, I.A.; Vilkovisky, G.A.
\newblock {Relativistic S Matrix of Dynamical Systems with Boson and Fermion
  Constraints}.
\newblock {\em Phys. Lett. B} {\bf 1977}, {\em 69},~309--312.
\newblock {\url{https://doi.org/10.1016/0370-2693(77)90553-6}}.

\bibitem[Hawking and Ellis(1973)]{hawking_ellis_1973}
Hawking, S.W.; Ellis, G.F.R.
\newblock {\em The Large Scale Structure of Space-Time}; Cambridge Monographs
  on Mathematical Physics; Cambridge University Press: {Cambridge, UK}, 1973.
\newblock {\url{https://doi.org/10.1017/CBO9780511524646}}.

\bibitem[Fathi and Jalalzadeh(2017)]{Fathi:2017pjm}
Fathi, M.; Jalalzadeh, S.
\newblock {Quantum Hamilton-Jacobi cosmology and classical-quantum
  correlation}.
\newblock {\em Int. J. Theor. Phys.} {\bf 2017}, {\em 56},~2167--2177.
\newblock {\url{https://doi.org/10.1007/s10773-017-3363-1}}.

\bibitem[Rashki and Jalalzadeh(2017)]{Rashki:2016udu}
Rashki, M.; Jalalzadeh, S.
\newblock {The Quantum State Of The Universe From Deformation Quantization and
  Classical-Quantum Correlation}.
\newblock {\em Gen. Rel. Grav.} {\bf 2017}, {\em 49},~14.
\newblock {\url{https://doi.org/10.1007/s10714-016-2178-3}}.

\bibitem[Jalalzadeh and Moniz(2014)]{Jalalzadeh:2014jka}
Jalalzadeh, S.; Moniz, P.V.
\newblock {Dirac observables and boundary proposals in quantum cosmology}.
\newblock {\em Phys. Rev. D} {\bf 2014}, {\em 89},~083504.
\newblock {\url{https://doi.org/10.1103/PhysRevD.89.083504}}.

\bibitem[Fathi et~al.(2016)Fathi, Jalalzadeh, and Moniz]{Fathi:2016lws}
Fathi, M.; Jalalzadeh, S.; Moniz, P.V.
\newblock {Classical Universe emerging from quantum cosmology without horizon
  and flatness problems}.
\newblock {\em Eur. Phys. J. C} {\bf 2016}, {\em 76},~527.
\newblock {\url{https://doi.org/10.1140/epjc/s10052-016-4373-5}}.

\bibitem[Rashki and Jalalzadeh(2015)]{Rashki:2014noa}
Rashki, M.; Jalalzadeh, S.
\newblock {Holography from quantum cosmology}.
\newblock {\em Phys. Rev. D} {\bf 2015}, {\em 91},~023501.
\newblock {\url{https://doi.org/10.1103/PhysRevD.91.023501}}.

\bibitem[Benedetti and Petronio(1992)]{Benedetti1992}
Benedetti, R.; Petronio, C.
\newblock {\em Lectures on Hyperbolic Geometry}; Springer: Berlin/Heidelberg, Germany, 1992.
\newblock {\url{https://doi.org/10.1007/978-3-642-58158-8_1}}.

\bibitem[Farooq and Ratra(2013)]{Farooq:2013hq}
Farooq, O.; Ratra, B.
\newblock {Hubble parameter measurement constraints on the cosmological
  deceleration-acceleration transition redshift}.
\newblock {\em Astrophys. J. Lett.} {\bf 2013}, {\em 766},~L7.
\newblock {\url{https://doi.org/10.1088/2041-8205/766/1/L7}}.

\bibitem[Scolnic et~al.(2018)]{Pan-STARRS1:2017jku}
Scolnic, D.M.;  et~al.
\newblock {The Complete Light-curve Sample of Spectroscopically Confirmed SNe
  Ia from Pan-STARRS1 and Cosmological Constraints from the Combined Pantheon
  Sample}.
\newblock {\em Astrophys. J.} {\bf 2018}, {\em 859},~101.
\newblock {\url{https://doi.org/10.3847/1538-4357/aab9bb}}.

\bibitem[Camarena and Marra(2018)]{Camarena:2018nbr}
Camarena, D.; Marra, V.
\newblock {Impact of the cosmic variance on $H_0$ on cosmological analyses}.
\newblock {\em Phys. Rev. D} {\bf 2018}, {\em 98},~023537.
\newblock {\url{https://doi.org/10.1103/PhysRevD.98.023537}}.

\bibitem[Hinshaw et~al.(2013)]{WMAP:2012nax}
Hinshaw, G.;  et~al.
\newblock {Nine-Year Wilkinson Microwave Anisotropy Probe (WMAP) Observations:
  Cosmological Parameter Results}.
\newblock {\em Astrophys. J. Suppl.} {\bf 2013}, {\em 208},~19.
\newblock {\url{https://doi.org/10.1088/0067-0049/208/2/19}}.

\bibitem[Chen et~al.(2019)Chen, Huang, and Wang]{Chen:2018dbv}
Chen, L.; Huang, Q.G.; Wang, K.
\newblock {Distance Priors from Planck Final Release}.
\newblock {\em JCAP} {\bf 2019}, {\em 2},~028.
\newblock {\url{https://doi.org/10.1088/1475-7516/2019/02/028}}.

\bibitem[Davari and Rahvar(2021)]{Davari:2021mge}
Davari, Z.; Rahvar, S.
\newblock {MOG cosmology without dark matter and the cosmological constant}.
\newblock {\em Mon. Not. Roy. Astron. Soc.} {\bf 2021}, {\em 507},~3387--3399.
\newblock {\url{https://doi.org/10.1093/mnras/stab2350}}.

\bibitem[Serra et~al.(2009)Serra, Cooray, Holz, Melchiorri, Pandolfi, and
  Sarkar]{Serra:2009yp}
Serra, P.; Cooray, A.; Holz, D.E.; Melchiorri, A.; Pandolfi, S.; Sarkar, D.
\newblock {No Evidence for Dark Energy Dynamics from a Global Analysis of
  Cosmological Data}.
\newblock {\em Phys. Rev. D} {\bf 2009}, {\em 80},~121302.
\newblock {\url{https://doi.org/10.1103/PhysRevD.80.121302}}.

\bibitem[Marra and Sapone(2018)]{Marra:2017pst}
Marra, V.; Sapone, D.
\newblock {Null tests of the standard model using the linear model formalism}.
\newblock {\em Phys. Rev. D} {\bf 2018}, {\em 97},~083510.
\newblock {\url{https://doi.org/10.1103/PhysRevD.97.083510}}.

\end{thebibliography}
\end{document}